\documentclass[trackchanges,twocolumn]{aastex701}

\usepackage{newtxtext,newtxmath}
\usepackage{tabularx}
\usepackage[T1]{fontenc}
\usepackage{graphicx}	
\usepackage{amsmath}	
\usepackage{threeparttable,booktabs}
\usepackage{float}
\usepackage{soul}
\usepackage{placeins}
\usepackage{atbegshi}
\usepackage{aecompl}
\usepackage{ulem}
\usepackage{xcolor}
\usepackage{CJK}
\usepackage{bm}
\usepackage{hyperref}

\newcommand{\msun}{\mathrm{M}_\odot}

\DeclareMathOperator{\arcsinh}{arcsinh}

\begin{document}
\begin{CJK*}{UTF8}{gbsn}

\title[Halo mass dependence on magnitude gap]{Dependence of halo properties on central-satellite magnitude gaps through weak lensing measurements}

\author[orcid=0009-0007-4996-685X]{Mingtao Yang (杨明焘)}
\affiliation{Department of Astronomy, Shanghai Jiao Tong University, Shanghai 200240, China}
\affiliation{State Key Laboratory of Dark Matter Physics, Key Laboratory for Particle Astrophysics and Cosmology (MOE), \& Shanghai Key Laboratory for Particle Physics and Cosmology, Shanghai Jiao Tong University, Shanghai 200240, China}
\email[hide]{jane550@sjtu.edu.cn}

\author[orcid=0000-0002-8010-6715]{Jiaxin Han}
\affiliation{Department of Astronomy, Shanghai Jiao Tong University, Shanghai 200240, China}
\affiliation{State Key Laboratory of Dark Matter Physics, Key Laboratory for Particle Astrophysics and Cosmology (MOE), \& Shanghai Key Laboratory for Particle Physics and Cosmology, Shanghai Jiao Tong University, Shanghai 200240, China}
\email[show]{jiaxin.han@sjtu.edu.cn} 

\author[orcid=0000-0002-5762-7571]{Wenting Wang}
\affiliation{Department of Astronomy, Shanghai Jiao Tong University, Shanghai 200240, China}
\affiliation{State Key Laboratory of Dark Matter Physics, Key Laboratory for Particle Astrophysics and Cosmology (MOE), \& Shanghai Key Laboratory for Particle Physics and Cosmology, Shanghai Jiao Tong University, Shanghai 200240, China}
\email[show]{wenting.wang@sjtu.edu.cn} 

\author[0000-0002-0610-2361]{Hekun Li}
\affiliation{Shanghai Astronomical Observatory, Chinese Academy of Sciences, 80 Nandan Road, Shanghai 200030, China}
\email{hekun_lee@shao.ac.cn}

\author[orcid=0009-0006-2694-6752]{Cong Liu} 
\affiliation{Department of Astronomy, Shanghai Jiao Tong University, Shanghai 200240, China}
\affiliation{State Key Laboratory of Dark Matter Physics, Key Laboratory for Particle Astrophysics and Cosmology (MOE), \& Shanghai Key Laboratory for Particle Physics and Cosmology, Shanghai Jiao Tong University, Shanghai 200240, China}
\email[hide]{lc-5351@sjtu.edu.cn@sjtu.edu.cn}

\author[orcid=0000-0003-0002-630X]{Jun Zhang} 
\affiliation{Department of Astronomy, Shanghai Jiao Tong University, Shanghai 200240, China}
\affiliation{State Key Laboratory of Dark Matter Physics, Key Laboratory for Particle Astrophysics and Cosmology (MOE), \& Shanghai Key Laboratory for Particle Physics and Cosmology, Shanghai Jiao Tong University, Shanghai 200240, China}
\email[hide]{betajzhang@sjtu.edu.cn}

\author[orcid=0000-0002-9767-9237]{Shuai Feng}
\affiliation{Hebei Key Laboratory of Photophysics Research and Application, College of Physics, Hebei Normal University, 20 South Erhuan Road, Shijiazhuang 050024, China}
\email{sfeng@hebtu.edu.cn}

\author[orcid=0000-0002-3073-5871]{Shiyin Shen}
\affiliation{Shanghai Astronomical Observatory, Chinese Academy of Sciences, 80 Nandan Road, Shanghai 200030, China}
\email[hide]{ssy.shao.ac.cn}

\author[orcid=0009-0000-3649-5365]{Zhenjie Liu} 
\affiliation{Department of Astronomy, Shanghai Jiao Tong University, Shanghai 200240, China}
\affiliation{State Key Laboratory of Dark Matter Physics, Key Laboratory for Particle Astrophysics and Cosmology (MOE), \& Shanghai Key Laboratory for Particle Physics and Cosmology, Shanghai Jiao Tong University, Shanghai 200240, China}
\affiliation{Division of Physics and Astrophysical Science, Graduate School of Science, Nagoya University, Nagoya 464-8602, Japan}
\email[hide]{liuzhj26@sjtu.edu.cn}

\author[orcid=0000-0003-3997-4606]{Xiaohu Yang} 
\affiliation{Department of Astronomy, Shanghai Jiao Tong University, Shanghai 200240, China}
\affiliation{State Key Laboratory of Dark Matter Physics, Key Laboratory for Particle Astrophysics and Cosmology (MOE), \& Shanghai Key Laboratory for Particle Physics and Cosmology, Shanghai Jiao Tong University, Shanghai 200240, China}
\email[hide]{xyang@sjtu.edu.cn}

\author{Yi Lu}
\affiliation{Shanghai Astronomical Observatory, Chinese Academy of Sciences, 80 Nandan Road, Shanghai 200030, China}
\email{luyi@shao.ac.cn}

\author[orcid=0000-0002-2986-2371]{Surhud More}
\affiliation{Inter-University Centre for Astronomy and Astrophysics, Post Bag 4 Ganeshkhind, Pune 411 007, India}
\email{surhud@iucaa.in}

\begin{abstract}

The magnitude gap between the central and satellite galaxies encodes information about the mass accretion history of a dark matter halo, and serves as a useful observational probe for the mass distribution in a halo. In this work, we perform the first weak lensing test of the connections between the magnitude gap and the halo profile. We measure the halo profiles of isolated central galaxies (ICGs) selected primarily from the SDSS Main Galaxy Sample. Halo mass and concentration are inferred by fitting stacked lensing profiles in bins of central luminosity, $L_\mathrm{c}$, and the central-satellite magnitude gap, $L_\mathrm{gap}$. We detect dependence on the magnitude gap in both halo properties. The dependence is the strongest in the ICG luminosity range of $10^{10.3}<L_\mathrm{c}[h^{-2}L_\odot]\leq 10^{10.7}$, where halos with smaller gaps have higher masses and lower concentrations. When $10^{10.7} <L_c[h^{-2}L_\odot] \leq 10^{11.1}$, however, no significant gap dependence is detected. In the range of $10^{9.9}<L_\mathrm{c}[h^{-2}L_\odot] \leq 10^{10.3}$, a disordering of the gap dependence is marginally observable. We compare the observational results with predictions by two lightcone catalogs built from the Illustris TNG300 and the Millennium simulations. The gap dependence in the two mock samples show overall consistency with observations, but neither matches them in all $L_\mathrm{c}$ bins to a quantitative level. We also compare the significance of the gap dependence on halo mass and concentration and find that our measurement prefers gap dependence in both parameters, while the halo mass dependence is preferred over the concentration if only one of the two dependencies is allowed. 
\end{abstract}


\section{Introduction}

In 1994, \cite{1994Natur.369..462P} discovered a bright elliptical galaxy, RXJ1340.6+4018, with an absolute $R$ band magnitude $M_R\approx -23.5$, embedded in an extended $x$-ray emitting hot gas halo. However, it is surrounded only by much fainter satellites. The authors suspect that the bright elliptical galaxy is relics of a merged galaxy group and thus call it a `fossil group' (FG). Later on, \cite{jones_multiwavelength_2000} measured the magnitude gap, $\Delta m_{12}=2.5$~mag, between this elliptical galaxy and its brightest satellite galaxy, indicating the brightest satellite is $10$ times fainter. Afterwards, \cite{jones_nature_2003} defined FGs observationally as the systems with $\Delta m_{12}\geq 2.0$, along with a lower limit on x-ray luminosity. This definition marked a milestone using the magnitude gap as an indicator of the halo's age.

The formation of such extreme systems as FGs, as well as the connection between the central-satellite magnitude gap and halo properties in a broader magnitude gap range can be understood from a dynamical perspective \citep{Deason_2013}. Satellite galaxies are stripped and disrupted by the tidal field of the host halo, gradually sinking into the halo center due to dynamical friction. More massive satellites are subject to stronger dynamical frictions \citep{chandrasekhar_dynamical_1943}, sinking to the host center at a faster rate, resulting in higher efficiencies in tidal stripping and disruption. When the brightest satellite eventually merges with the central galaxy, the gap of the system is enlarged. It can be narrowed by the subsequent accretion of new satellites that are brighter than the current second brightest galaxy (SBG) or further enlarged by the infall of fainter satellite galaxies. Old galaxy systems that formed earlier have largely ceased accreting new satellites and experienced more internal mergers, thus on average have larger magnitude gaps. On the contrary, young galaxy systems that are more actively accreting at later times have continuous replenishment of new satellites to keep the magnitude gaps small. The magnitude gap, in such a way, is related to the merger history of a halo, or, the age of a halo. In this work, we define the magnitude gap, which is 2.5 times the luminosity gap, as the magnitude difference between the central galaxy and the SBG in a galaxy system, unless otherwise stated. This is expressed as: 
\begin{equation}
    \Delta m = m_\mathrm{s}-m_\mathrm{c} = 2.5\log (L_\mathrm{c}/L_\mathrm{s}),
\end{equation}
where $m_\mathrm{c}$ and $m_\mathrm{s}$ are the absolute magnitudes, and $L_\mathrm{c}$ and $L_\mathrm{s}$ are the absolute luminosities of the central galaxy and the SBG, respectively. 

Among the various interesting directions that can be explored about the magnitude gap, one practical and interesting topic is to use it to help better constrain the host halo properties~\citep[see e.g.][]{2018ARA&A..56..435W}. The traditional proxies of the host halo mass are descriptions of the brightness or richness in the current state of a halo, such as the stellar mass or luminosity of the central galaxy, or the number of galaxies brighter than certain threshold in a  galaxy clusters/groups. As mentioned above, the magnitude gap is an imprint of the halo merger history, thus provides extra information than the traditional proxies, constraining the histories to reach the current state brightness or richness. A further merit of the magnitude gap is that it is relatively easy to derive from photometric surveys, making it an economic observable for probing the properties of the host halo.

A series of studies explore the magnitude gap dependence on host halo properties in the context of the galaxy luminosity function. \cite{paranjape_luminosities_2012} proposed that the gap distribution originates from discrete sampling of the luminosity function, and discussed whether a global or conditional luminosity function (CLF) is required. The latter got clear answers from \citet{more_magnitude_2012, hearin_mind_2013} and \cite{shen_statistical_2014} that the CLF is required, and order statistics from the CLF is capable of interpreting the observational luminosity distributions of the group galaxies and the distribution of the magnitude gap, revealing the statistical origin of the magnitude gap dependence on host halo mass. As a followup exploitation, \cite{2015ApJ...804...55L} empirically calibrates the halo mass - gap relations based on CLF mocks. A more recent study by \cite{zhou_mining_2022} incorporates the magnitude gaps between the central galaxy and satellite galaxies in different ranks of luminosity as proxies to the shapes of different segments of the CLF, and study the gap dependence in the context of the CLF.

To establish the connection between halo properties and the magnitude gap in observations, it is essential to first derive the halo properties through observables. A few studies \citep[e.g.][]{zarattini_fossil_2015, 2021MNRAS.500.3776W} use the satellite abundance as a proxy to the halo mass. They measured the satellite luminosity functions of galaxy systems with different magnitude gaps, and reported that systems with larger such gaps have less satellites, indicating galaxy systems with larger magnitude gaps are on average hosted by less massive halos. The studies by \cite{hearin_mind_2013} and \cite{golden-marx_impact_2018} estimate the halo mass from the velocity dispersion of satellites. It is found that the magnitude gap provides constraints on both the mass-richness relation and the stellar to halo mass relation (SHMR). Specifically, large-gap groups exhibit higher halo masses at fixed richness but lower halo masses at fixed stellar mass of the brightest central galaxy (BCG) compared with small-gap groups. 

As one of the most direct methods for measuring the halo mass profile, however, weak lensing constraints on the gap dependence of halo properties are still absent. Thus in this study, we perform detailed analysis on the quantitative dependence of halo profile parameters on the magnitude gaps, with the halo mass profiles directly measured through weak lensing.

The layout of this paper is as follows. In Section \ref{sec:data}, we describe the galaxy samples used, including the foreground lens sample and the background source sample. The simulation data to be compared against our observational measurements is also introduced in this section. In section \ref{sec:method}, we enumerate the methods utilized in our lensing measurement, from the catalog construction, to the excess surface density (ESD) measurements, and finally to the determination of halo properties. In Section~\ref{sec:data analysis}, we analyze the data by first showing the selection effects on our lens sample and our binning strategy, then presenting the measured ESD profiles, and finally presenting our weak lensing result of the dependence of host halo properties on the magnitude gaps, after calibrating the bias of the best-fit halo properties. In Section~\ref{sec:Fitting results and comparisons against simulations}, we compare the weak lensing measurement with predictions by the Illustris TNG300 and Millennium simulations. We discuss the significance of dependence on the magnitude gap between the halo mass and the concentration in Section~\ref{sec:significance} and conclude in Section \ref{sec:conclusion}.

\section{Data}
\label{sec:data}
\subsection{Isolated central galaxies and their satellites}\label{sec:ICGs}

In order to select a sample of galaxies that are highly likely the true central galaxies of their host dark matter halos, we select the so-called isolated central galaxies following similar selections as \cite{2021MNRAS.500.3776W, 2021ApJ...919...25W,2023ApJ...947...19A,2025arXiv250303317W}.
The parent galaxy sample for selection is a combination of the Main Galaxy Sample (MGS) in the seventh data release of the Sloan Digital Sky Survey (SDSS/DR7; \cite{2009ApJS..182..543A}) and a complementary sample that supplements a large amount of spectral redshifts for the galaxies that are not observed in SDSS/DR7 due to fiber collisions (\cite{feng_bivariate_2019}). The complementary sample includes galaxies from SDSS/DR18 and several other spectroscopic surveys including LAMOST \citep[]{luo_first_2015} and GAMA \citep[]{baldry_galaxy_2018}. The readers can refer to \cite{2016RAA....16...43S} for more details. Combining these two samples, we achieve around $95\%$ completeness of the galaxies above the $r$-band flux limit of $17.77$.

The isolated central galaxies (ICGs) are selected as the brightest galaxies within the projected virial radius of their host dark matter halos and within three times the virial velocity in the line-of-sight direction. The virial radius and velocity are estimated from stellar mass using the abundance matching method in \cite{Guo10}. In this way, we have a preliminary ICG catalog, which has 556,980 ICGs. 

While we include additional data to the SDSS/DR7 spectroscopic data to account for galaxies without spectroscopic observations, still 5\% of the galaxies are not spectroscopically observed in later surveys.
This may result in some preliminarily selected ICGs having potentially brighter companions that are not included in the spectroscopic sample. To avoid such contamination, we further use the photometric redshift (photo-z) probability distribution \citep{2009MNRAS.396.2379C} to identify potentially brighter photometric companions lacking a spectroscopic redshift. We discard ICGs that have brighter photometric companions within its projected virial radius. Only companions that have a greater than 10\% probability in sharing the ICG redshift are considered in this process. By the end of this stage, we have 536,177 ICGs. 

To eliminate cases where the ICG candidate is locally brightest within its small virial radius but is within the virial radius of a larger halo, hence is in fact a satellite galaxy, we require the ICGs must not be projected within the virial radius and within three times the virial velocity along the line of sight of another more massive galaxy. After this step, we have 482,889 ICGs selected. Tested with a mock galaxy catalog based on the semi-analytical model of \cite{2010MNRAS.404.1111G}, the completeness of ICGs among all true halo central galaxies at this stage is about 90\%. The purity is above 82\%, which reaches $>$90\% at $\log_{10}M_{\ast,\mathrm{ICG}}/\msun>11.5$. For further details, readers may refer to \cite{2019MNRAS.487.1580W}. 

The magnitude gaps are calculated between each ICG and the SBG projected within the virial radius and within three times the virial velocity along the line of sight. For precise measurement of the magnitude gap, the SBG is also required to have a spectroscopic redshift (spec-z) measurement. Systems with the photometrically identified SBG brighter than the spectroscopically confirmed SBG are excluded. In other words, we only use systems in which the SBG is brighter than the flux limit of SDSS Main galaxies ($r\sim17.77$). After applying this cut, our ICG catalog contains 54,569 ICGs in total. 

For the weak lensing analysis, we only use ICGs within the footprint of the background source catalog (see Section~\ref{sec:shearcat}) and within the central luminosity and magnitude gap bins described in Section~\ref{subsec:selection effect}, resulting in a final number of 30,649 ICGs. 

The host halo masses of these ICGs are measured through weak lensing, and in the following subsections, we introduce the background sources and the shear catalog used for the weak lensing analysis.

\subsection{the shear catalog and background sources}
\label{sec:shearcat}

Our shear catalog includes the multiple moments of the projected 2D power spectrum of galaxy image in Fourier space, $G_1,G_2,N,U,V$, for around 116 million source galaxies. They can be understood as counterparts of `ellipticities' in real space in traditional shear estimations \citep[see][]{2017ApJ...834....8Z}. We will introduce the definitions of these multiple moments and the calculation of them in Section~\ref{sec:fourierquad} below.

The background source sample for galaxy-galaxy (g-g) lensing measurements is drawn from the data of DECam Legacy Survey (DECaLS) \cite{2019AJ....157..168D}, a photometric survey primarily designed to support the target selection of the follow-up spectroscopic surveys on the Dark Energy Spectroscopic Instrument (DESI) \citep{2016arXiv161100036D}. DECaLS provides optical imaging in the $g$, $r$ and $z$ bands, covering approximately 10,000 deg$^2$. For g-g lensing measurements, we use the $z$ band data. The redshifts for the galaxies are obtained from the photo-z catalog of \cite{zou_photometric_2019}, where redshifts are estimated using a linear regression relation between the spectroscopic redshifts and the photometric spectral energy distribution (SED). The typical photo-z uncertainty is $\sigma_z\simeq 0.017(1+z)$. 

In our measurement, the sample of ICGs serve as foreground lenses. We only use ICGs with a redshift greater than $0.03$. This is because for lenses with very low redshifts, the lensing efficiency is low, which would increase the noise in the measured lensing signals. The background sources for each ICGs are required to have their photometric redshifts higher than the ICG by $0.2$ to reduce contamination from foreground galaxies due to the photo-z uncertainty. 

\subsection{the Illustris TNG300 simulation}
\label{sec:TNG300}

The IllustrisTNG simulations are a suite of hydrodynamical simulations carried out with the moving-mesh code \citep[\textsc{arepo};][]{Springel2010}. Comprehensive treatments of various galaxy formation and evolution processes are incorporated, including metal line cooling, star formation and evolution, chemical enrichment, and gas recycling. The TNG suites of simulations adopt the Planck 2015 $\Lambda$CDM cosmological model with $\Omega_\mathrm{m}=0.3089$, $\Omega_\Lambda=0.6911$, $\Omega_\mathrm{b}=0.0486$, $\sigma_8=0.8159$, $n_\mathrm{s}=0.9667$, and $h=0.6774$ \citep{Planck2015}. For more details about TNG, we refer the readers to \cite{Marinacci2018,Naiman2018,Nelson2018,Pillepich2018,Springel2018,Nelson2019}.

The sets of TNG simulations cover different box sizes and mass resolutions. In this paper, we will use the TNG300-1 (hereafter TNG300) simulation, which has the largest boxsize among all runs as well as the highest mass resolution at this boxsize. TNG300 has a periodic comoving box with 205~$h^{-1}\mathrm{Mpc}$ on each side that follows the joint evolution of 2,500$^3$ dark matter particles and $\sim$2,500$^3$ gas cells. Each dark matter particle has a mass of $4.0\times10^7 \,h^{-1}\msun$, while the baryonic mass resolution is $7.6\times10^6 \,h^{-1}\msun$. 

In the analysis of systematics in Section~\ref{sec:model sel eff}, we will model the observational selection effects by constructing a mock sample from discrete snapshots of TNG300. And to analyze the estimator bias of the halo mass and concentration estimated from lensing signals with respect to the mean value in each bin, 
we carry out Monte Carlo (MC) simulations, with the fiducial halo mass and halo concentration distribution taken from TNG300. In the end, we perform detailed comparisons between our measurements of the mass - gap relation in real observation and predictions from the TNG300 and Millennium simulations in Section~\ref{sec:compare with simulations}. 

\subsection{the Millennium simulation}\label{sec:Millennium}
The Millennium Simulation is a cosmological N-body simulation performed with Gadget \citep{2005Natur.435..629S}. The simulation models the evolution of $2160^3$ particles with particle mass $8.6\times 10^8 \,h^{-1}M_{\odot}$ within a periodic box length of 500 $h^{-1}\textrm{Mpc}$. The gravitational growth is traced by these particles from $z=127$ to $0$ in a $\Lambda \mathrm{CDM}$ cosmology ($\Omega_m = 0.25$, $\Omega_{\Lambda}=0.75$, $h=0.73$, $n=1$, $\sigma_8=0.9$ ) most consistent with the Wilkinson Microwave Anisotropy Probe (WMAP) year 1 data \citep{spergel_first-year_2003}. 

In order to understand the baryonic content of the model universe, semi-analytical modelling (SAM) \citep[e.g.][]{1991ApJ...379...52W,1993MNRAS.264..201K,1994MNRAS.271..781C,1999MNRAS.303..188K,1999MNRAS.310.1087S,2000MNRAS.311..576K,2001MNRAS.320..504S,2003MNRAS.343...75H,2005ApJ...631...21K,2005MNRAS.364..407C,2006MNRAS.370.1651C,2007MNRAS.375.1189M,2007MNRAS.375....2D,2011MNRAS.413..101G, 2012MNRAS.423.1992S} is applied on dark matter halo merger trees to reveal key baryonic processes, including gas cooling, star formation, reionization heating, supernova feedback, mergers, black hole growth, metal enrichment and feedback from active galactic nuclei.

To assess the galaxy formation model variations, we will compare the mass - gap relation derived from g-g lensing observations with those from the TNG300 and Millennium simulations in Section~\ref{sec:compare with simulations}. A lightcone mimics the geometric and photometric effects in real sky surveys, thus enabling a direct comparison between simulated and observational results. We make use of the Henriques2012a all-sky lightcone \citep[see][]{henriques_confronting_2012,overzier_millennium_2013} in the Millennium database for this purpose. The recipe for baryonic physics is taken from the SAM in \cite{guo_dwarf_2011} and the multi-wavelength properties of galaxies are derived using the stellar population synthesis model in \cite{bruzual_stellar_2003}. The redshift extends to $z=4.35$, far beyond the redshift limit of around $0.4$ in the SDSS Main Galaxy Survey. With a flux limit of $r<17.77$ added, we can obtain a counterpart of the observed sample. 

\section{Method}
\label{sec:method}
\subsection{Fourier Quad method for building the shear catalog}
\label{sec:fourierquad}

The shear catalogue introduced in Section~\ref{sec:shearcat} is constructed with the Fourier Quad (FQ) method \citep{10.1111/j.1365-2966.2007.12585.x, 2010MNRAS.403..673Z, 2011JCAP...11..041Z, 2011MNRAS.414.1047Z, 2015JCAP...01..024Z, 2017ApJ...834....8Z}. \cite{2019ApJ...875...48Z} and \cite{Zhang_2022} constructed the FQ pipeline that converts raw images to the shear catalog. The process includes background removal, identification of cosmic ray and other image defects, astrometric calibration, source/noise identification, super-flat field correction, and calculation of the shear estimators in Fourier space. 

FQ calculates the following quantities from the 2D power spectrum of each source, which are subsequently used to construct estimators of the shear signal, 
\begin{equation}
G_1 = -\frac{1}{2}\int{d^2\bm{k}(k^2_x-k^2_y)T(\bm{k})M(\bm{k})},
\label{eq_shear_1}
\end{equation}
\begin{equation}
G_2 = -\int{d^2\bm{k}k_x k_y T(\bm{k})M(\bm{k})},
\label{eq_shear_2}
\end{equation}
\begin{equation}
N = \int{d^2\bm{k}\left[k^2-\frac{\beta^2}{2}k^4\right]T(\bm{k})M(\bm{k})},
\end{equation}
where $\bm{k}$ is the wave vector, and
\begin{equation}
T(\bm{k}) = \left|\tilde{W}_{\beta}(\bm{k}) \right|^2 /  \left|\tilde{W}_{PSF}(\bm{k}) \right|^2,
\end{equation}
\begin{equation}
M(\bm{k}) = \left|\tilde{f}^S(\bm{k}) \right|^2 - \tilde{F}^S -  \left|\tilde{f}^B(\bm{k}) \right|^2 + \tilde{F}^B.
\end{equation}
$T(\bm{k})$ represents the ratio between the power spectrum of an isotropic Gaussian function, $\tilde{W}_{\beta}(\bm{k})$, and the power spectrum of the original PSF. $\tilde{W}_{\beta}(\bm{k})$ is defined as:
\begin{equation}
    \tilde{W}_{\beta}(\bm{x}) = \frac{1}{2\pi \beta^2} \mathrm{exp}\left(-\frac{\left|\bm{x} \right|^2}{2\beta^2} \right),
\end{equation}
where $\beta$ is the scale radius of the Gaussian function.
$T(\bm{k})$ transforms the form of the original PSF to the desired Gaussian form in order to correct for the PSF effect. $M(\bm{k})$ corrects the power spectrum of the source by subtracting the contributions of the background and the Poisson noise. $\tilde{f}^S(\bm{k})$ and $\tilde{f}^B(\bm{k})$ are the Fourier transformations of the galaxy and the background noise, respectively. $\tilde{F}^\mathrm{S}$ and $\tilde{F}^\mathrm{B}$ are estimates of the power spectrum of the Poisson noise for the galaxy image and the background, respectively.

It can be shown that the ensemble averages of $G_\mathrm{1}$ and $G_\mathrm{2}$ satisfy the following relations \citep{2011MNRAS.414.1047Z} :
\begin{equation}
    \frac{\left<G_1 \right>}{\left<N \right>}=g_1+O(g_{1,2}^3), \frac{\left<G_2 \right>}{\left<N \right>}=g_2+O(g_{1,2}^3),
\end{equation}
where $g_1$ and $g_2$ are the two shear components. In a more recent work, \citet{2017ApJ...834....8Z} proposed a new approach for the shear measurement, called the PDF Symmetrization Method (PDF-SYM). PDF-SYM symmetrizes the probability distribution function (PDF) of $G_1-\hat{g}_1\left(N+U \right)$ and $G_2-\hat{g}_2\left(N-U \right)$ in order to determine the shear values $\hat{g}_1$ and $\hat{g}_2$. This introduces two additional terms ($U$ and $V$) to the three initially defined ($G_1$, $G_2$, and $N$). The new terms are defined as:

\begin{equation}
\label{Ud}
U = -\frac{\beta^2}{2}\int{d^2\bm{k} (k_x^4-6k_x^2k_y^2+k_y^4)T(\bm{k})M(\bm{k})},
\end{equation}
\begin{equation}
\label{Vd}
V = -2\beta^2 \int{d^2\bm{k} (k_x^3k_y-k_xk_y^3)T(\bm{k})M(\bm{k})}.
\end{equation}
$U$ is introduced to guarantee that the distribution of $G_\mathrm{1}-\hat{g_\mathrm{1}}(N+U)$ and $G_\mathrm{2}-\hat{g_\mathrm{2}}(N-U)$ are symmetric around 0, whereas $V$ is needed to transform $U$ when a coordinate rotation appears in the shear measurement. \citet{2017ApJ...834....8Z} proved that PDF-SYM allowed the shear estimation to reach the lowest theoretical statistical limit (Cramer-Rao limit). 

FQ has been applied to the images in all three bands of  DECaLS and in $i'$ band of CFHTLenS , with field distortion tests to evaluate the quality of shear recovery. All bands in the two surveys show equally well results and are recommended for utilization except for the DECaLS $g$ band, which might be due to some frequency dependent systematics of the instruments. It is shown that the quality of shear recovery by the FQ pipeline does not depend on the seeing conditions of the images \citep[see Figure~ 16 of ][]{Zhang_2022}. The DECaLS images with relatively poor seeing conditions, that has the typical Full-Width-at-Half-Maximum (FWHM) of the point spread function (PSF) about 1.4-1.6 arcsec, still yield accurate shear estimators.

In this work, we use the shear catalog computed from the DECaLS $z$-band photometries, which has a lower resolution but a three times larger sample size than CFHTLenS. According to \cite{shen_tele-correlation_2025}, the shear bias depends on the redshift cut applied to the source galaxy sample. Because the redshift distribution for our ICG bins vary (see Section~\ref{subsec:selection effect} and Figure~\ref{fig:sel_eff}), and the redshifts of source galaxies are required to be larger than that of the lens by at least 0.2, the source galaxies behind each ICG population exhibits different redshift distribution. We have performed field distortion tests on the source galaxies behind each lens population using the same methodology as in Appendix B of \cite{liu_ellipticities_2025}. No significant bias was detected. Therefore, we do not apply any corrections to our shear catalog.

\subsection{PDF Symmetrization method to obtain the excess surface density profiles} \label{subsec:ESD data}

For a spherically symmetric lens, the tangential shear $\gamma_\mathrm{t}$, which measures the shape distortion of a source galaxy along the radial and tangential directions to the projected lens (in our case ICGs), is related to the excess surface density (ESD) of the halo at a projected radius $R$ as
\begin{equation}
    \Delta\Sigma(R)\equiv \Sigma(<R)-\Sigma(R)=\gamma_\mathrm{t}(R)\Sigma_\mathrm{crit},\label{eq:lensing eqn}
\end{equation}
where $\Sigma(<R)$ is the mean surface density within the radius $R$, and $\Sigma(R)$ is the surface density at $R$. The critical density is defined as
\begin{equation}
    \Sigma_\mathrm{crit}\equiv \frac{c^2}{4\pi G}\frac{D_\mathrm{s}}{D_\mathrm{ls}D_\mathrm{l}},\label{eq:sigma_crit}
\end{equation} where $D_\mathrm{s},D_\mathrm{l},D_\mathrm{ls}$ are the angular diameter distance of the source, the lens, and that between the two, respectively~\citep[see e.g.][]{1991ApJ...370....1M,1994ApJ...437...56F,2005astro.ph..9252S}. For aspherical lenses,  Equation~\eqref{eq:lensing eqn} still holds for azimuthally averaged measurements.

When measuring the ESD profile, the physical radial range from $10^{-3}$ to $10\,h^{-1}\mathrm{Mpc}$, centered on the lens in the projected plane, is divided into 20 equal bins in logarithmic space. In each bin, we estimate the ESD using the PDF-SYM method, which determines the value that optimally symmetrizes the PDF of  
\begin{equation}
\epsilon=G_\mathrm{t}-(\Delta\Sigma_\mathrm{PDF-SYM}/\Sigma_\mathrm{crit})(N+U_\mathrm{t}).
\end{equation}
Here, $G_\mathrm{t}$ and $U_\mathrm{t}$ are derived through a coordinate rotation, expressed as follows,
\begin{align}
    G_\mathrm{t} &= G_1 \cos(2\theta)-G_2 \sin(2\theta),\\
    U_\mathrm{t} &=U \cos(4\theta)-V \sin(4\theta),
\end{align} where $\theta$ is the clockwise angle between the RA direction and the line perpendicular to the projected connection between the lens and the source.

We split bins over the lens sample according to the black grids shown in Figure~\ref{fig:sel_eff} and measure the ESD profile for each bin. The covariance matrix of the ESD measurements from different radial bins are estimated using the jackknife method in which the sky coverage of the lens sample is divided into $90$ patches.

\subsection{Profile fitting for halo properties}
\label{sec:spf}

For each $(L_\mathrm{c},L_\mathrm{gap})$ bin, we fit a Navarro-Frenk-White (NFW) profile \citep{1997ApJ...490..493N} to the observed ESD profile by maximizing the likelihood
\begin{align}
\begin{split}
    \mathcal{L}=& \frac{1}{(2\pi)^{d/2}|C|^{1/2}}\\
    & \exp[-\frac{1}{2}(\hat{\bm{\Delta\Sigma}}-\bm{\Delta\Sigma}_\mathrm{PDF-SYM})^\top C^{-1}(\hat{\bm{\Delta\Sigma}}-\bm{\Delta\Sigma}_\mathrm{PDF-SYM})]\label{eq:L-spf},
\end{split}
\end{align}
where $d$ is the number of radial bins used for fitting and $C$ is the covariance matrix. $\Delta\Sigma_\mathrm{PDF-SYM}$ is the measured ESD profile and $\hat{\Delta\Sigma}$ is the ESD profile predicted from the NFW model as follows \citep[see][]{1997ApJ...490..493N,1996A&A...313..697B,2000ApJ...534...34W},
\begin{align}
    \hat{\Delta\Sigma}(R) &= 2R_\mathrm{s} \rho_0 \left[ f_\mathrm{p}\left(\frac{R}{R_\mathrm{s}}\right)-f_{\rm c}\left(\frac{R}{R_{\rm s}}\right)\right],\label{eq:NFW_sigma} \\[10pt]
    \rho_0 &= \frac{M_\mathrm{200b}}{4\pi R_\mathrm{s}^3 \left[\ln(1+c_\mathrm{200b}) - c_\mathrm{200b}/(1+c_\mathrm{200b})\right]}, \\[10pt]
    f_\mathrm{p}(x) &=
    \begin{cases}
        \frac{1}{x^2-1} + \frac{\arcsinh\left(\sqrt{1-x^2}/x\right)}{(1-x^2)^{1.5}}, & \text{if } x < 1, \\[10pt]
        \frac{1}{3}, & \text{if } x = 1, \\[10pt]
        \frac{1}{x^2-1} - \frac{\arcsin\left(\sqrt{x^2-1}/x\right)}{(x^2-1)^{1.5}}, & \text{if } x > 1,
    \end{cases} \\[10pt]
    f_\mathrm{c}(x) &= \frac{2f(x)}{x^2}, \\[10pt]
    f(x) &=
    \begin{cases}
        \ln(0.5x) + \frac{\arcsinh\left(\sqrt{1-x^2}/x\right)}{\sqrt{1-x^2}}, & \text{if } x < 1, \\[10pt]
        1 - \ln 2, & \text{if } x = 1, \\[10pt]
        \ln(0.5x) + \frac{\arcsin\left(\sqrt{x^2-1}/x\right)}{\sqrt{x^2-1}}, & \text{if } x > 1,\label{eq:NFW_sigma_end}
    \end{cases}
\end{align} 
where $R_\mathrm{s}$ is the scale radius. In this study, the halo mass, $M_\mathrm{200b}$, is the total mass within $R_\mathrm{200b}$, with $R_\mathrm{200b}$ defined as the virial radius within which the mean density equals $200$ times the background (matter) density of the universe. Defining the concentration parameter as $c_\mathrm{200b}=R_\mathrm{200b}/R_\mathrm{s}$, the mass and concentration parameters fully specify an NFW profile.

Throughout this paper, we use $\hat{M}$ and $\hat{c}$ to denote the best recovered $M_\mathrm{200b}$ and $c_\mathrm{200b}$ from the lensing ESD profiles.\footnote{In the simulation-based context, $M_\mathrm{200b}$ and $c_\mathrm{200b}$ are directly taken from the friends-of-friends(FoF) halo group catalog.} The covariance matrix of the derived mass and concentration is approximated by inverting the Hessian matrix of the negative log-likelihood function, evaluated at the best-fit location.

To ensure the robustness of these fits, we carefully select the radial range for the analysis. In the center of a halo, baryons become more dominant and may cause the deviation from the NFW model profile adopted above. Moreover, the ICGs may be offset from the actual potential minimum, which may cause the measured central density profiles centered on ICGs more flattened. At large distances, density contributions from neighbouring halos become important and the profile starts to deviate from Equation~\eqref{eq:NFW_sigma}. Therefore, we need to choose a radial range that avoids the very central and outer parts. Explicitly, we choose the radial range (in projection) to be from $0.06R_\mathrm{200b}$ to $2.3R_\mathrm{200b}$. The choice of $0.06R_\mathrm{200b}$ follows \cite{2015MNRAS.446.1356H}, whereas the choice of $2.3R_\mathrm{200b}$ follows the depletion radius definition for halo boundary \citep[e.g.][]{2021MNRAS.503.4250F,2022MNRAS.513.4754F,2023ApJ...953...37G}, which is shown to be an optimal halo exclusion radius \citep{2023MNRAS.525.2489Z,2025ApJ...979...55Z}. Here we determine the values of $R_\mathrm{200b}$ for different $L_\mathrm{c}$ and $L_\mathrm{gap}$ bins according to the mean virial radius of central galaxies in the Illustris TNG300 simulation. In Table~\ref{tab:R200b}, we provide the values of $R_\mathrm{200b}$ adopted for our $L_\mathrm{c}$ and $L_\mathrm{gap}$ bins. 

\begin{table*}
    \centering
    \caption{$R_\mathrm{200b}$ used for scaling when determining the radial range for fitting an NFW profile to the measured ESD profile. These radii are the  mean $R_\mathrm{200b}$ in each $L_\mathrm{c}$ and $L_\mathrm{gap}$ bin extracted from the Illustris TNG300 simulation at $z=0$.}
\label{tab:R200b}    
    \begin{tabular}{cccc}
    \hline\hline
      \multicolumn{4}{c}{$R_\mathrm{200b}[h^{-1}\mathrm{Mpc}]$}\\\hline
         & 
     $9.9<\textrm{log}L_\mathrm{c}[{h}^{-2}L_{\odot}]\leq 10.3$&$10.3<\textrm{log}L_\mathrm{c}[{h}^{-2}L_{\odot}]\leq 10.7$&$10.7<\textrm{log}L_\mathrm{c}[{h}^{-2}L_{\odot}]\leq 11.1$\\
 $0<\textrm{log}L_\mathrm{gap}\leq 0.3$& $0.399$&$0.620$&$0.953$\\
 $0.3<\textrm{log}L_\mathrm{gap}\leq 0.6$& $0.369$&$0.580$&$0.922$\\
 $0.6<\textrm{log}L_\mathrm{gap}\leq 1$& $0.338$&             $0.537$&...\\
 \hline 
 \end{tabular}
\end{table*}

In Appendix~\ref{sec:test radial choice}, we test the influences of different choices of radial ranges on the best-fit halo mass, and find that our results are insensitive to the exact boundary choices over a reasonable range.

\section{Data analysis}\label{sec:data analysis}

\subsection{Selection function and binning}\label{subsec:selection effect}

\begin{figure}
    \centering
    \includegraphics[width=\linewidth]{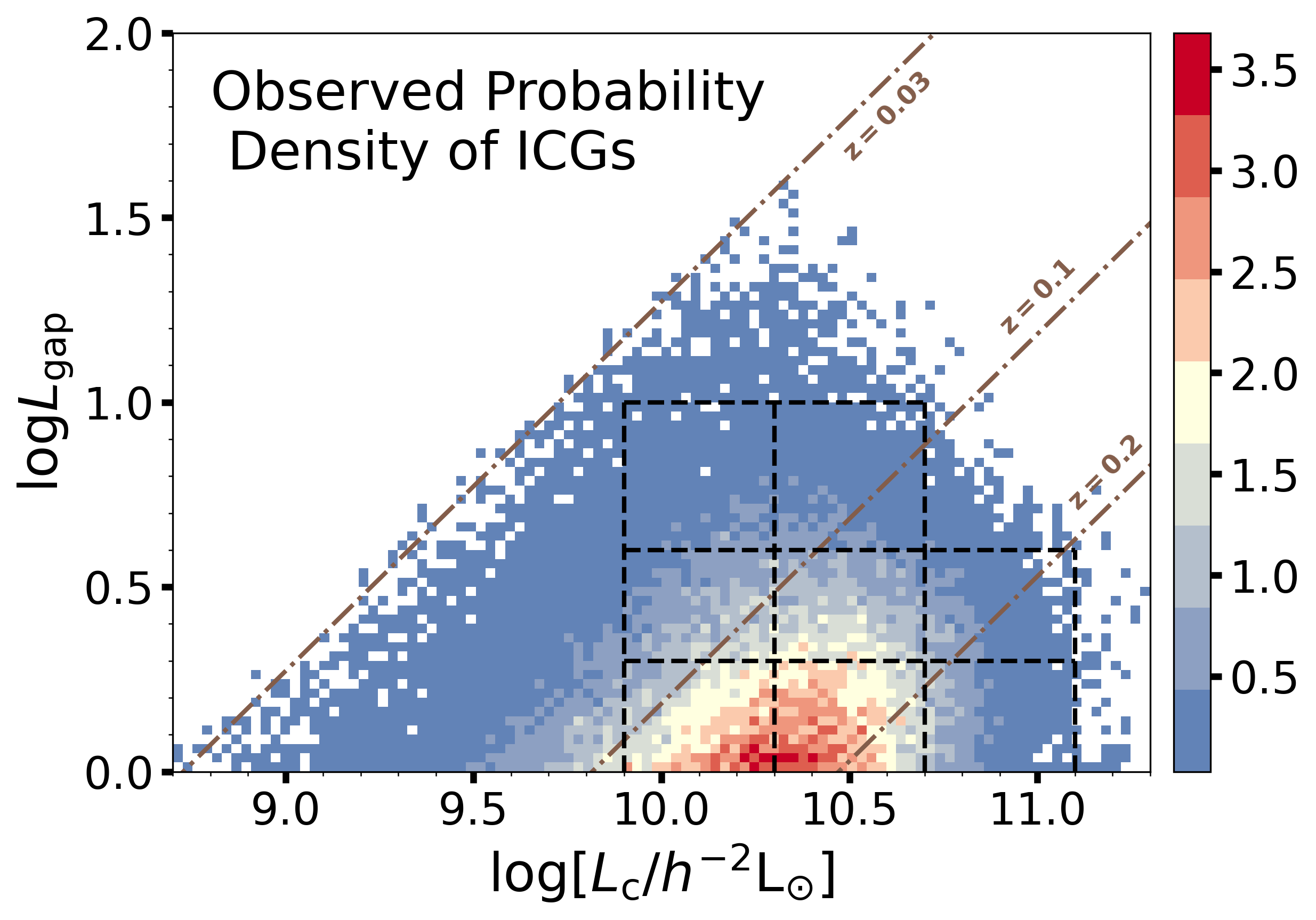}
    \caption{Distribution of the selected ICGs in the $L_\mathrm{c},L_\mathrm{gap}$ plane, where $L_\mathrm{c}$ represents the luminosity of the ICG and $L_\mathrm{gap}$ denotes the luminosity ratio between the ICG and its brightest satellite galaxy (or the SBG of the system). The color bar indicates the probability density of ICGs. Brown dash-dotted lines show the boundaries above which the SBGs are too faint to be observed at the corresponding redshift as labelled. Black dashed lines mark the boundaries of the eight $L_\mathrm{c},L_\mathrm{gap}$ bins used for stacked lensing analysis.}
    \label{fig:sel_eff}
\end{figure}

In this study, we require both the ICG and SBG to have spec-z measurements. Thus, whether a system can be observed depends on whether the apparent magnitude of the SBG is above the flux limit of $17.77$ of the SDSS spectroscopic Main galaxy sample. In Figure~\ref{fig:sel_eff}, we show the distribution for our sample of ICGs over the plane of $\log L_\mathrm{gap}$ versus $\log L_\mathrm{c}$. Throughout this paper, we use $L_\mathrm{c}$  and $L_\mathrm{s}$ to denote the absolute luminosity of the central and the second brightest galaxy (SBG), and $L_\mathrm{gap}$ is defined through $\textrm{log}L_\mathrm{gap}=\textrm{log}(L_\mathrm{c}/L_\mathrm{s})$. The brown dash-dotted lines show the boundaries above which SBGs are too faint to be observed at redshifts of $z=0.03$, $z=0.1$ and $z=0.2$, respectively. As the redshift increases, only systems with a brighter SBG, and hence a smaller gap, are observable. 

\begin{figure}
    \centering
    \includegraphics[width=\linewidth]{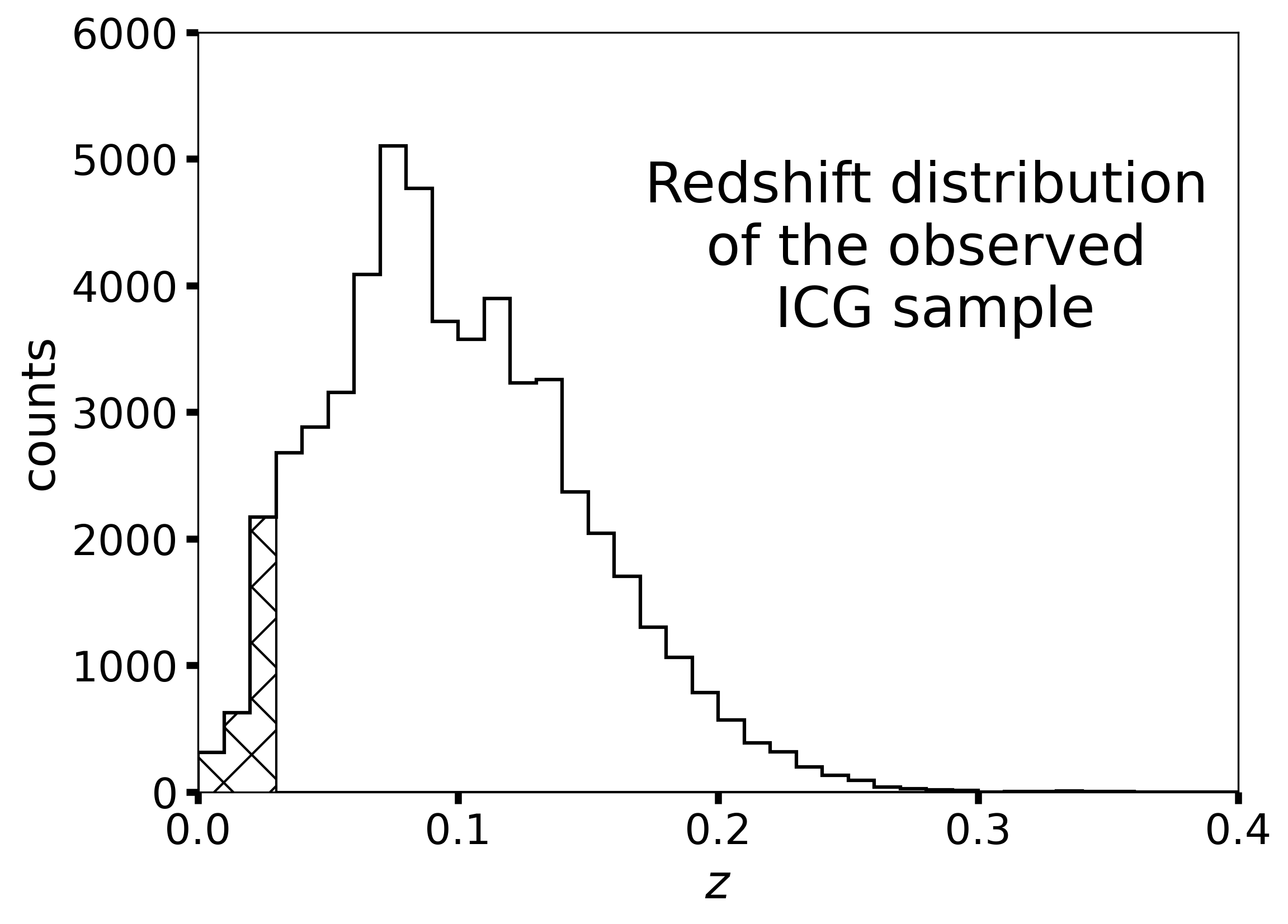}
    \caption{The redshift distribution of the observed ICG sample. The hatched region indicates that ICGs with $z\leq 0.03$ are removed from the lens sample for higher lensing efficiency.}
    \label{fig:obs_z_dist}
\end{figure}

The redshift distribution of the observational ICG sample is shown in Figure~\ref{fig:obs_z_dist}. Over $99\%$ of the sample have redshifts smaller than 0.25.

In our analysis, we choose eight bins. The bin edges for $\textrm{log}L_\mathrm{c}$ and $\textrm{log}L_\mathrm{gap}$ are $9.9,10.3,10.7,11.1$, and $0,0.3,0.6,1$, respectively. The eight bins are demonstrated by the black dashed-line grids in Figure~\ref{fig:sel_eff}. 
 
\subsection{ESD measurements}

\begin{figure*}
    \centering
    \includegraphics[width=0.9\textwidth]{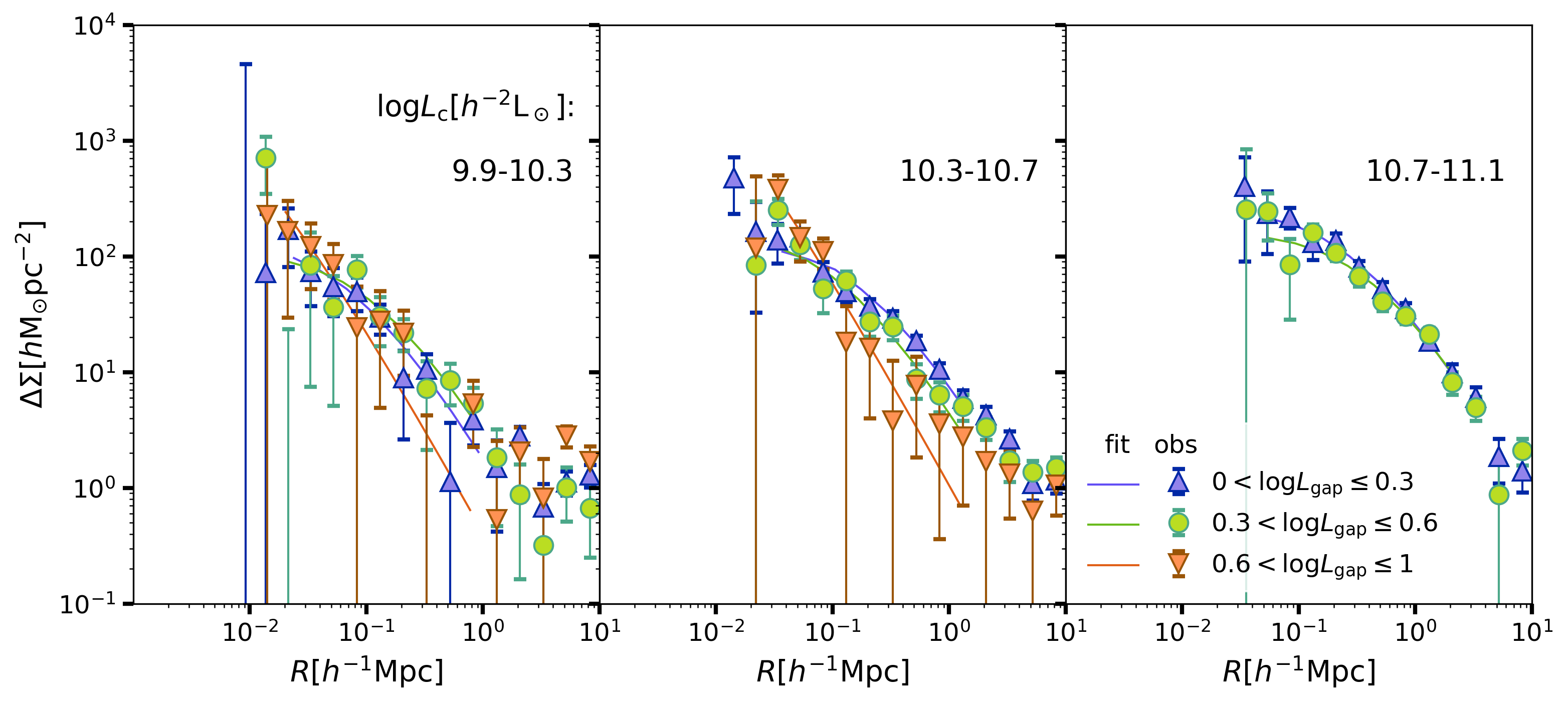}
    \caption{The measured lensing ESD profiles centered on ICGs in different bins of $L_\mathrm{c}$ (see the text in each panel) and $L_\mathrm{gap}$ (different color symbols in the same panel, see the legend). The bins are defined in Figure~\ref{fig:sel_eff}. The $x$-axis is the projected physical radius to the ICG. The symbols and associated errorbars show the estimated ESDs with the PDF-SYM method and their $1\sigma$ uncertainty estimated from 90 jackknife subsamples. The solid lines with corresponding colors show the best-fit ESD profiles within $[0.06-2.3R_\mathrm{200b}]$ based on the NFW model profile (see Section~\ref{sec:spf}).} 
    \label{fig:ESD-data}
\end{figure*}

The measured excess surface density (ESD) profiles are shown in Figure~\ref{fig:ESD-data}.  Each panel corresponds to a given bin in $\log L_\mathrm{c}$ (see the text in each panel), and within a given panel, symbols with different colors are measured lensing ESD profiles centered on ICGs with different $\log L_\mathrm{gap}$.

In the right and middle panels, we can see some weak trend that for ICGs with smaller $\log L_\mathrm{gap}$, their lensing ESD profiles are slightly higher in amplitudes, indicating that there are some weak dependence of the host halo profiles on magnitude gaps. Measurements in the left panel is quite noisy. No clear dependence on $\log L_\mathrm{gap}$ is shown.

We obtain the best-fit or maximum likelihood (ML) halo mass (log$\hat{M}$) and concentration ($\log \hat{c}$) following the approach described in Section~\ref{sec:spf}. The best-fit projected NFW profiles are plotted as solid curves with corresponding colors indicating different gap bins in Figure~\ref{fig:ESD-data}. Note we only plot the best-fit projected NFW model profiles over the projected radial range where the actual data points are used for the fitting. According to the best fits, we can see that the best-fit halo profiles show dependencies on the magnitude gaps. 

\subsection{Calibration of estimator bias}
\label{sec: model est bias}
\begin{figure}
    \centering
    \includegraphics[width=\linewidth]{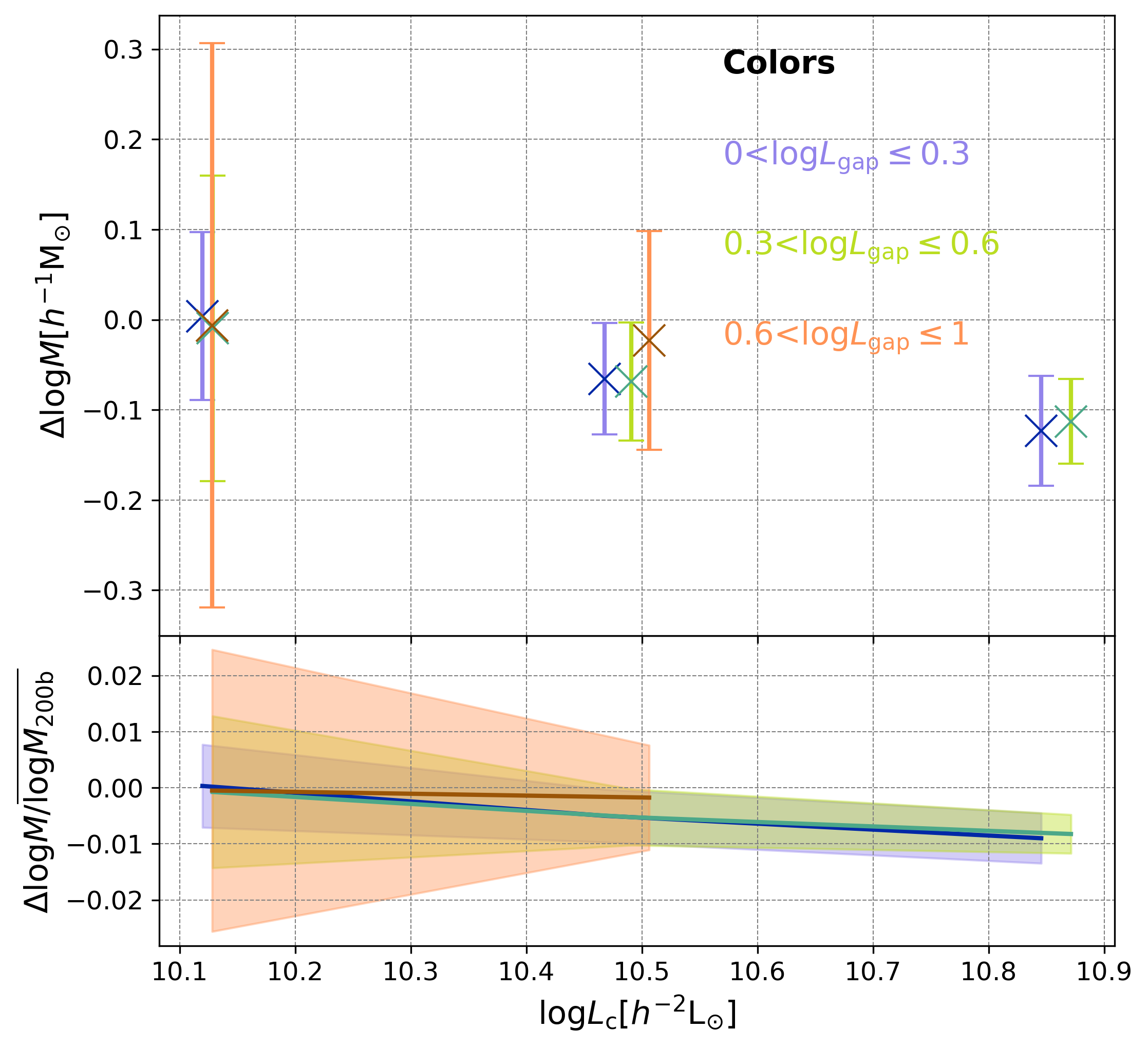}
    \caption{Top: The bias of the lensing estimated $\log \hat{M}$ from the mean $\log M_\mathrm{200b}$ in the lightcone sample. Crosses with error bars show the average and $1\sigma$ scatter of $\Delta \log M$ in 100 lightcone samples in each bin of $L_\mathrm{c}$ and $L_\mathrm{gap}$. Bottom: The above bias relative to the mean $\log M_\mathrm{200b}$ in the lightcone sample. The solid lines together with the shaded regions show the mean and $1\sigma$ scatter of the relative bias in 100 lightcone samples. Colors represent different bins of $L_\mathrm{gap}$, as labeled in the legend.}
    \label{fig:logM_est_bias}
\end{figure}
\begin{figure}
    \centering
    \includegraphics[width=\linewidth]{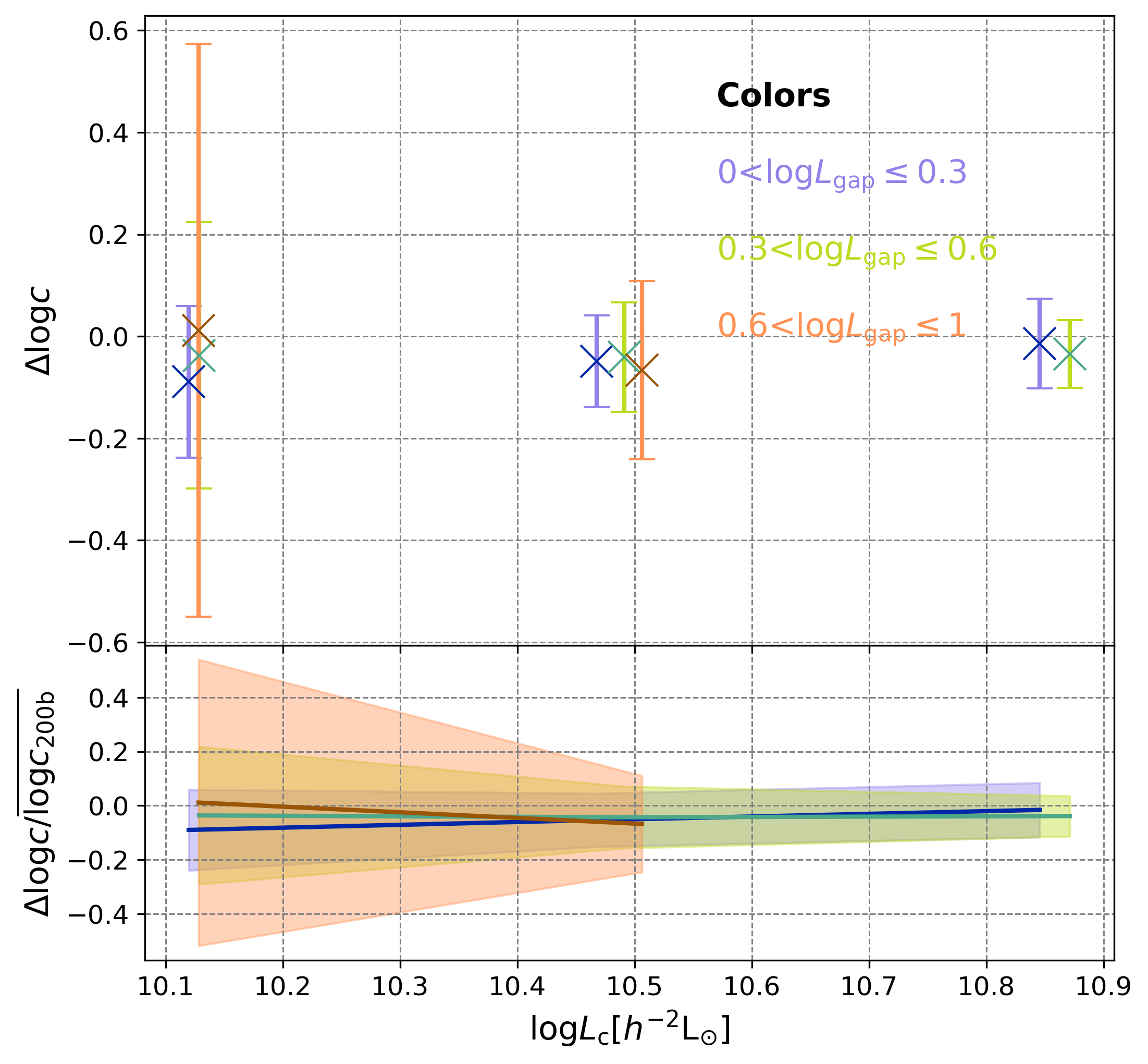}
    \caption{Same as Figure~\ref{fig:logM_est_bias}, but shows the lensing estimator bias of $\log \hat{c}$. }
    \label{fig:logc_est_bias}
\end{figure}

In each galaxy bin, the measured halo profiles are the average profiles for a sample of halos with different mass and concentration parameters. As we fit a single NFW profile to the stacked lensing profile, the best-fit halo parameters are not necessarily the same as the average parameters for the halos in the bin which can be more directly compared against theoretical predictions. To address this, we calibrate these best-fit values to the mean values of $\log M_\mathrm{200b}$ and $\log c_\mathrm{200b}$ in each $L_\mathrm{c}$ and $L_\mathrm{gap}$ bin.

The bias is modeled using Monte Carlo simulations of the lensing analysis procedures. We generate Monte Carlo realizations of the lensing system by sampling galaxies from the TNG300 simulation and assigning them to the observed galaxy positions, and repeat the lensing analysis on the Monte Carlo realizations to assess the estimator bias. The detailed steps are introduced in the following.

\begin{itemize}
\item \textit{Generate the lens sample:} To create a lens sample with known halo properties that mimics the redshift and luminosity distribution of the real lens sample, we generate a random lightcone sample of ICGs from TNG300 snapshots. The lightcone sample is a composition of sub-samples in very fine redshift bins. Each sub-sample is selected at the snapshot closest to the redshift under the flux limit of $r<17.77$, and is matched in size as the real lens sample in the corresponding redshift bin. The steps in building the lightcone sample are described in more detail in Section~\ref{sec:model sel eff} when we model the selection effects on the TNG300 data. Each lens from the lightcone sample is then randomly assigned to the position of one ICG from the observational sample in the same fine redshift bin. This way a mock lens sample with similar galaxy properties and known halo properties is generated at the same positions as the real observations. 

\item \textit{Randomize source shapes:} We rotate each background galaxy in the real observation by a random angle $\theta$ between $0$ to $2\pi$, to generate a random realization of the intrinsic shape of a source galaxy. The multiple moments of the power spectrum of the galaxy image in Fourier space, $G_1,G_2,N, U,V$, are updated after the rotation as follows.
\begin{equation}
    \begin{aligned}
        G'_1&=G_1\cos(2\theta)-G_2\sin(2\theta)\\
        G'_2&=G_1\sin(2\theta)+G_2\cos(2\theta)\\
         N'&=N\\
        U'&=U\cos(4\theta)-V\sin(4\theta)\\
        V'&=U\sin(4\theta)+V\cos(4\theta),
    \end{aligned}
\end{equation}
where the primed variables indicate the quantities after rotation. In this way, we smear out the shear signals generated by real lenses while the statistics of the shape noise in the real observation are still maintained. 

    \item \textit{Add shear signals:} With the lenses and sources generated above, shear signals are further added to the source shapes assuming the NFW density profiles for the lens halos. The reduced tangential shear signal, $g_\mathrm{t}\equiv \gamma_\mathrm{t}/(1-\kappa)=\Delta\Sigma/(\Sigma_\mathrm{crit}-\Sigma)$, can be analytically derived under the NFW assumption (see Equation~\ref{eq:NFW_sigma} to~\ref{eq:NFW_sigma_end}). The tangential shear signal is transformed into the RA and DEC coordinate system and added to the initialized $G_1'$, $G_2'$ in the previous step through
\begin{equation}
            \begin{aligned}
                G''_1 &= G_1'+g_1(N'+U')\\
                G''_2 &=G_2'+g_2(N'-U'),
            \end{aligned}
\end{equation}
where $G''_1,G''_2$ are the updated multiple moments of the power spectrum of galaxy image with the generated shear added.

\item \textit{Apply the lensing estimator:} We derive the stacked ESD profiles for the lightcone lens sample in each bin of $L_\mathrm{c}$ and $L_\mathrm{gap}$ given the shear catalog obtained above using the PDF-SYM method, and obtain the best-fit halo properties following the same steps as for the real observations.

Repeating these procedures for 100 times with different random realizations, we obtain 100 estimates of $\Delta\log M$ and $\Delta\log c$, which are defined as the differences between the lensing estimated $\log \hat{M}$ and $\log \hat{c}$ and the mean of $\log M_\mathrm{200b}$ and $\log c_\mathrm{200b}$ in the TNG lightcone sample. These estimator biases are displayed in Figure~\ref{fig:logM_est_bias} and~\ref{fig:logc_est_bias}. In most bins of $L_\mathrm{c}$ and $L_\mathrm{gap}$, both halo mass and concentration of the lensing estimation are biased low. However, the average bias of $\log \hat{M}$ and $\log \hat{c}$ relative to the sample mean throughout the 100 lightcone samples are within 1\% and 10\%, respectively. The relatively larger scatter in the largest $L_\mathrm{gap}$ bins is due to the larger uncertainty of lensing estimation in these bins. 

We calibrate the lensing-estimated $\log \hat{M}$ and $\log \hat{c}$ in observations to estimate the mean $\log M_\mathrm{200b}$ and $\log c_\mathrm{200b}$ within each $L_\mathrm{c}$-$L_\mathrm{gap}$ bin. This calibration involves subtracting the bias estimator, $\langle\Delta \log M\rangle_{100}$ and $\langle\Delta \log c\rangle_{100}$, which are calculated as the average estimator bias over 100 MC realizations, from the lensing best-fit values, $\log \hat{M}$ and $\log \hat{c}$, respectively. The uncertainty of the bias-calibrated estimator is determined as the geometric mean of the uncertainties of the lensing best fit and the bias estimator. The calibration process is expressed mathematically as follows:
\begin{equation}
    \log\hat{M}_\mathrm{bc} = \log \hat{M}-\langle\Delta \log M\rangle_{100}\label{eq:M_bc}
,\end{equation}
\begin{equation}
    \log\hat{c}_\mathrm{bc} = \log \hat{c}-\langle\Delta \log c\rangle_{100}\label{eq:c_bc}
,\end{equation}
\begin{equation}
    \sigma^2(\log\hat{M}_\mathrm{bc})=\sigma^2(\log\hat{M})+\sigma^2(\Delta \log M)/100,\label{eq:var_M_bc}
\end{equation}
\begin{equation}
    \sigma^2(\log\hat{c}_\mathrm{bc})=\sigma^2(\log\hat{c})+\sigma^2(\Delta \log c)/100,\label{eq:var_c_bc}
\end{equation}
where the subscript `$\mathrm{bc}$' indicates the bias-calibrated estimator. The comparisons between the lensing best fits and the mean estimators are shown in Figures~\ref{fig:logM_before_n_after_calibration}
 and~\ref{fig:logc_before_n_after_calibration}, for halo mass and concentration, respectively.  The minimal difference between the results before and after bias calibration indicates the best fit halo properties from the stacked ESD profiles in each bin of $L_\mathrm{c}$ and $L_\mathrm{gap}$ are unbiased estimators of the mean halo properties within the bin.

\begin{figure}
    \centering
    \includegraphics[width=\linewidth]{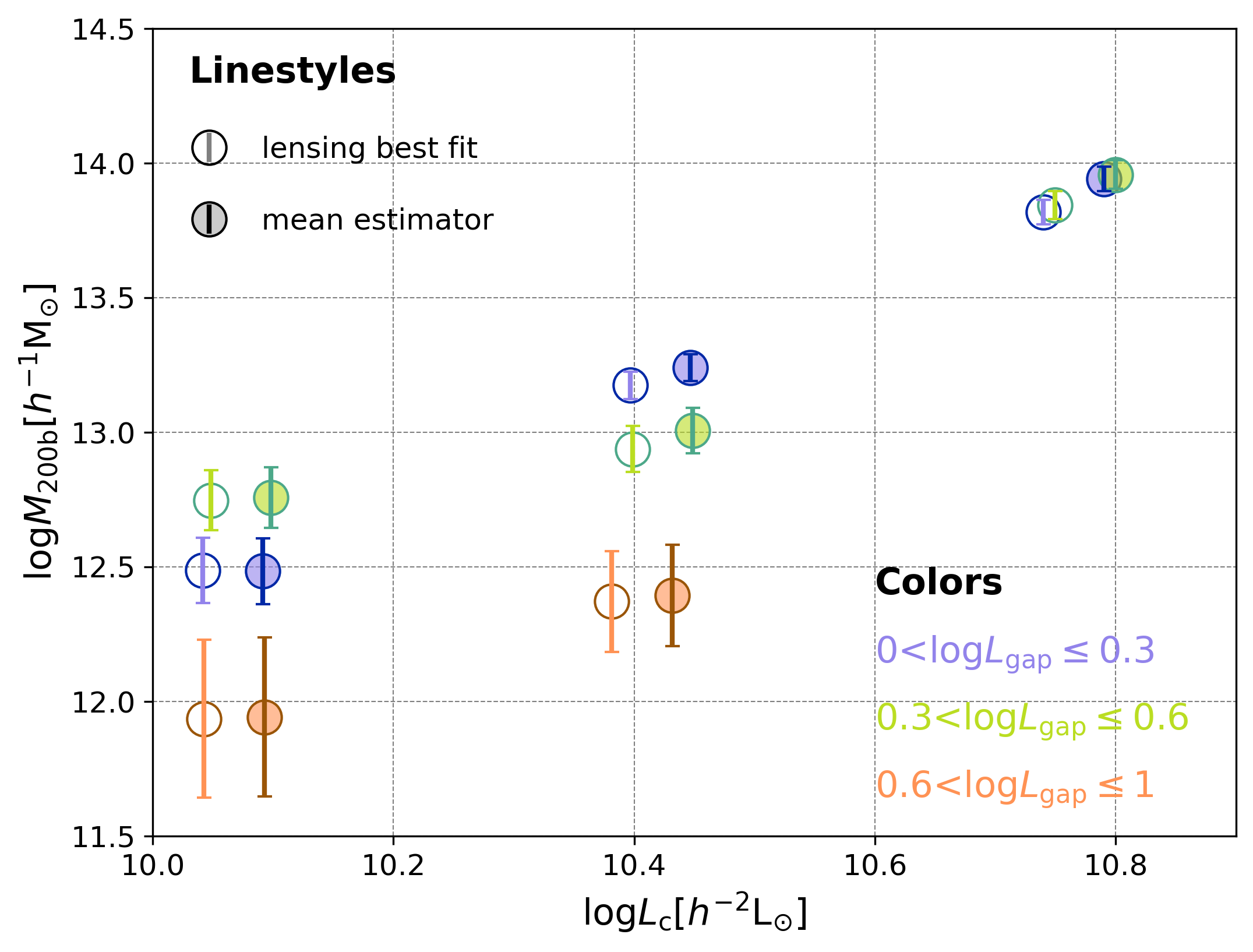}
    \caption{Best fit halo mass ($\log \hat{M}$) from measured ESD profiles shown by the unfilled circles, compared with the bias-calibrated mean estimator of $\log M_\mathrm{200b}$, $\log \hat{M}_\mathrm{bc}$, marked by filled circles in each bin of $L_\mathrm{c}$ and $L_\mathrm{gap}$. The errorbars of the unfilled circles represent the $1\sigma$ uncertainty of the lensing best fit. The errorbars of the filled circles demonstrate the $1\sigma$ uncertainty of the mean estimator, which is a combination of uncertainties from both the lensing best fit and the bias estimator. The $x$ values show the mean $\log L_\mathrm{c}$ in each bin. For visual clarity, the lensing best fits are shifted left by 0.05 dex.}
    \label{fig:logM_before_n_after_calibration}
\end{figure}
\begin{figure}
    \centering
    \includegraphics[width=\linewidth]{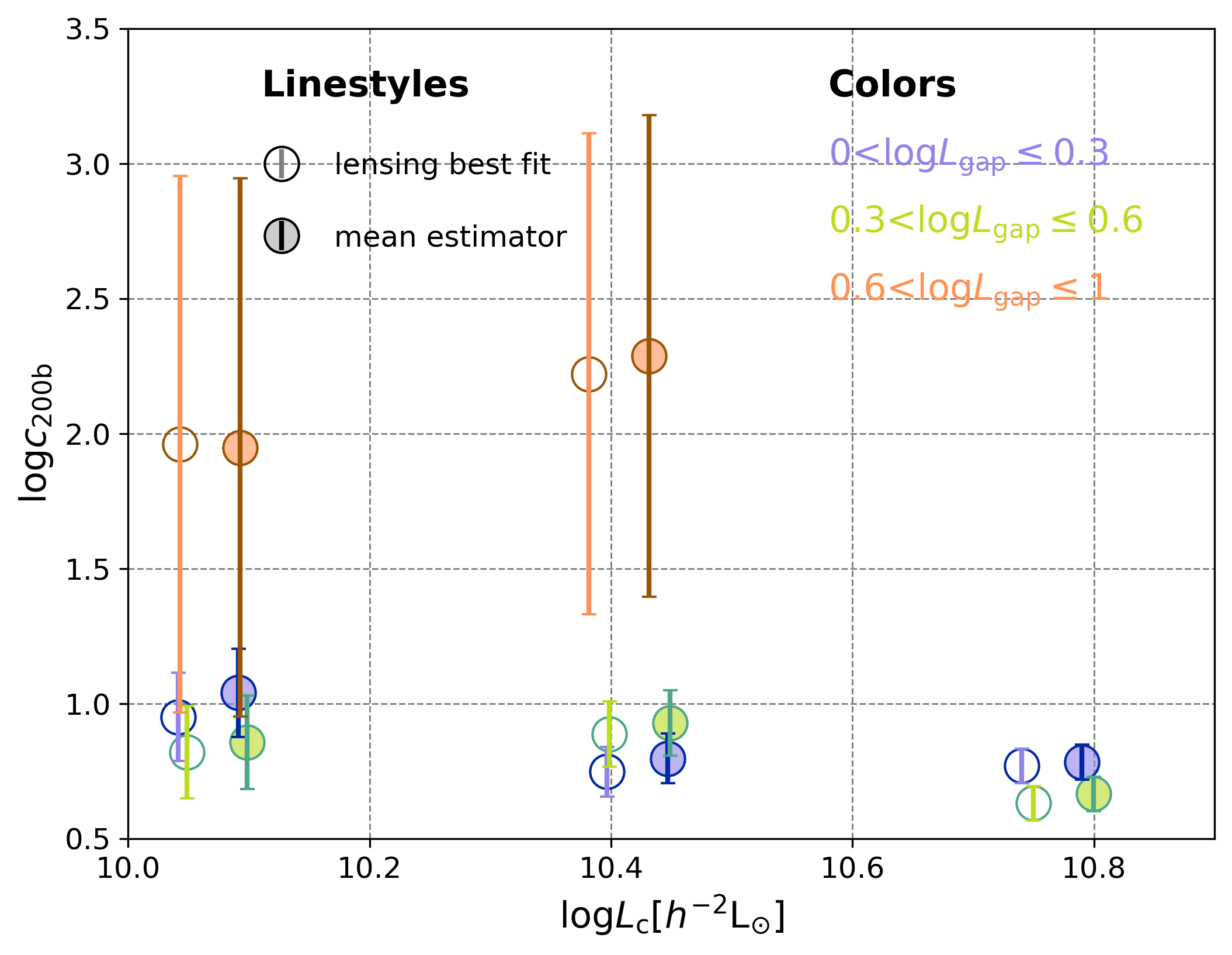}
    \caption{Similar to Figure~\ref{fig:logM_before_n_after_calibration}, but showing the results for concentration.}
    \label{fig:logc_before_n_after_calibration}
\end{figure}

\end{itemize}
\subsection{Dependence of host halo mass and concentration on magnitude gaps}\label{sec:Mh_c_gap}

Although the bias calibration has minimal impact on the final results, it simplifies the interpretation of the observational results and ensures a more equitable comparison with the simulated mean halo properties. For this reason, in the following analysis and in later comparisons against simulations, we focus on the bias-calibrated results.

As seen from filled circles in Figure~\ref{fig:logM_before_n_after_calibration}, the gap dependence of halo mass is most pronounced in the intermediate $L_\mathrm{c}$ bin. In this range, halo mass decreases monotonically with increasing $L_\mathrm{gap}$, with mean halo mass differences between neighboring gap bins exceeding the $1\sigma$ error margins. At the faint $L_\mathrm{c}$ end, larger error bars reduce the significance of the differences between gap bins. Additionally, the mean halo masses for the smallest two gap bins exhibit a reversed order compared with the intermediate bin, showing slightly separated errorbars. At the bright $L_\mathrm{c}$ end, no significant dependence on $L_\mathrm{gap}$ is observed, as the mean halo masses and the error bars of the two gap bins almost entirely overlap.

The halo concentration - gap dependence shown in Figure~\ref{fig:logc_before_n_after_calibration} resembles the halo mass - gap dependence, being most pronounced in the intermediate $L_\mathrm{c}$ bin. In this range, the concentration increases monotonically with $L_\mathrm{gap}$, and the errorbars for the largest and intermediate gap bins are clearly separated. At the faint $L_\mathrm{c}$ end, the constraints of gap dependence is poor given the tangled errorbars of the three gap bins. At the bright $L_\mathrm{c}$ end, the relative ordering of concentration for the two gap bins reverses relative to what is shown in the intermediate $L_\mathrm{c}$ bin, but this reversal remains marginally consistent within the error bars. 

\begin{figure}
    \centering
    \includegraphics[width=\linewidth]{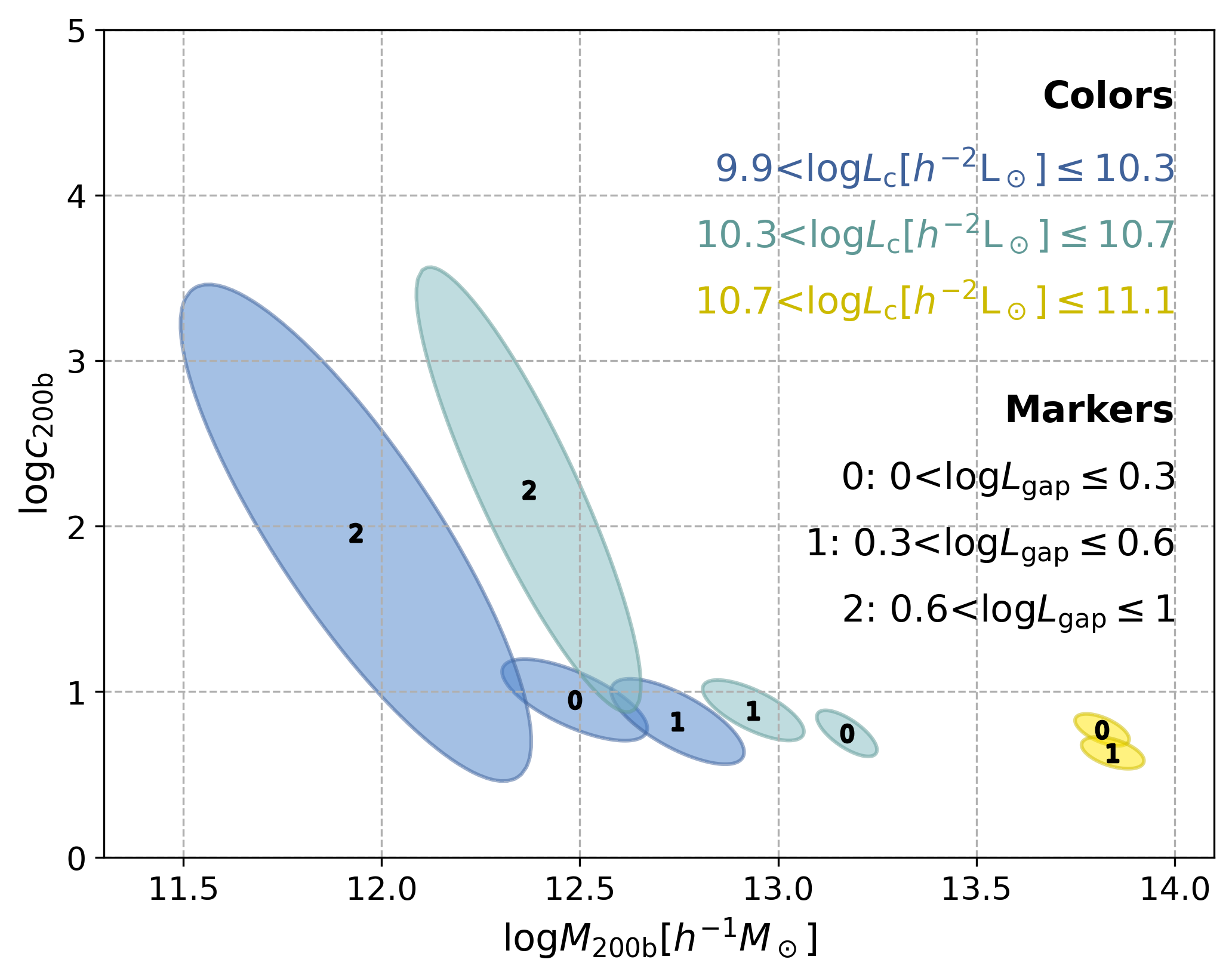}
    \caption{The $68.3\%$ confidence regions of the best-fit mass ($\log M_\mathrm{200b}$) and concentration ($\log c_\mathrm{200b}$) parameters from the weak lensing measurement. The confidence region is plotted according to the covariance matrix derived from the maximum likelihood estimation, assuming the posterior distribution as a joint Gaussian distribution. As indicated by the legend, colors denote different $L_\mathrm{c}$ bins, while different gap bins are distinguished by the marker numbers. }
    \label{fig:M-c-contour}
\end{figure}

Note that the errorbars shown in Figures~\ref{fig:logM_before_n_after_calibration} and \ref{fig:logc_before_n_after_calibration} are the marginalized errors on each parameter, while the two parameters are fit jointly. When examined in the 2-dimensional parameter space, the significance of the reversal becomes even lower. As shown in Figure~\ref{fig:M-c-contour}, the 68.3\% confidence regions for different gap bins partly overlap with each other at the lowest and highest central luminosities, respectively. On the other hand, the confidence regions at the intermediate luminosity are still separate from each other, showing the highest significance.

The observed gap dependence is largely consistent with previous theoretical understandings~\citep[e.g.,][]{Deason_2013}. For galaxy systems with larger magnitude gaps at fixed $L_\mathrm{c}$, the mass accretion is inactive. Satellites falling in earlier eventually merge with the central galaxy to increase $L_\mathrm{c}$ and increase the magnitude gaps, while there are less newly accreted satellites to decrease the magnitude gaps. The halo mass of such systems grows less efficiently at later times and are thus small. Such systems also have higher concentrations, as they accumulated most of their mass in a relatively early phase of the universe when the background matter density was higher. 

In the brightest $L_\mathrm{c}$ bin, the absence of a clear gap dependence may be understood by considering the mass to light ratio variation of galaxies. The brightest galaxies reside in the most massive halos with the highest total mass to light ratios. At the same time, the mass to light ratio drops quickly as the luminosity decreases, so that a small magnitude difference corresponds to a large difference in halo mass. As a result, when a central galaxy in the brightest bin merges with a satellite of lower luminosity, the increment in halo mass is relatively small.  Towards the low luminosity end, the decrease of mass to light ratio with galaxy luminosity slows down and reaches a minimum at a halo mass of $\sim 10^{12}M_\odot$ ~\citep[e.g.,][]{yang_galaxy_2008,Yang12,2015MNRAS.446.1356H}, below which it starts to increase with decreasing luminosity. This trend means that the response of halo mass increment to satellite merger at a given luminosity ratio is the smallest at the bright end, consistent with what we observe in Figure~\ref{fig:logM_before_n_after_calibration}.

\section{Comparisons against simulations}\label{sec:Fitting results and comparisons against simulations}

\subsection{Forward modeling selection effects with lightcone mocks} \label{sec:model sel eff}
\begin{figure}
    \centering
    \includegraphics[width=\linewidth]{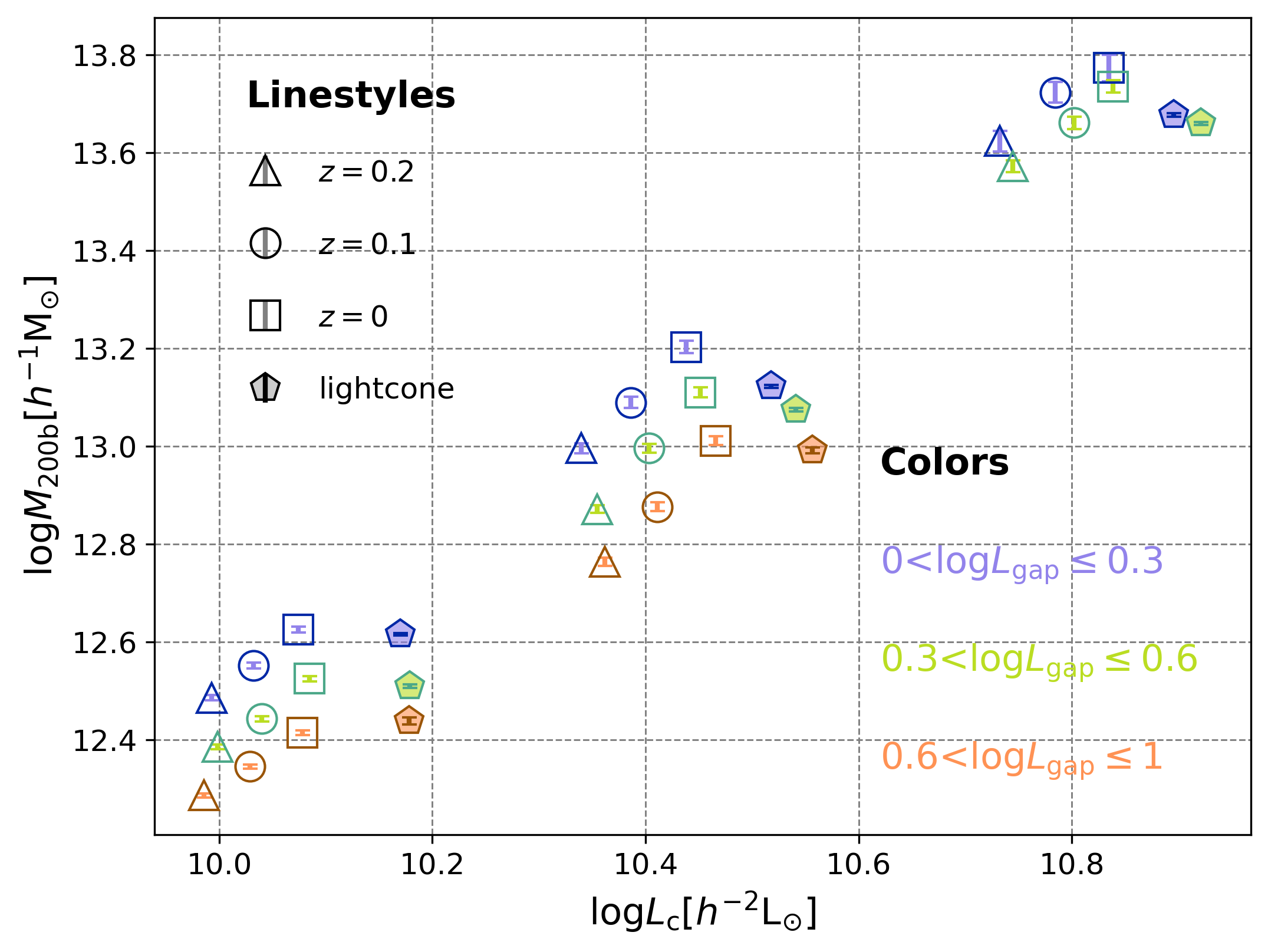}
    \caption{The impact of selection effects on the halo mass - gap relation. Triangles, circles, and squares illustrate the mean $\log M_\mathrm{200b}$ versus the mean $\log L_\mathrm{c}$ in each bin of $L_\mathrm{c}$ and $L_\mathrm{gap}$ at snapshots of $z=0$, 0.1, and 0.2, respectively. Error bars show the $1\sigma$ uncertainty of the sample mean. Pentagons with error bars demonstrate the average and $1\sigma$ scatter of the mean $\log M_\mathrm{200b}$ in each bin of $L_\mathrm{c}$ and $L_\mathrm{gap}$ in 100 lightcone samples. Colors represent different bins of $L_\mathrm{gap}$, as labeled in the legend. For visual clarity, the $x$ values of triangles, circles, and pentagons are shifted 0.1 dex to the left, 0.05 dex to the left, and 0.05 dex to the right.}
    \label{fig:logM_snapshots_lightcone}
\end{figure}
\begin{figure}
    \centering
    \includegraphics[width=\linewidth]{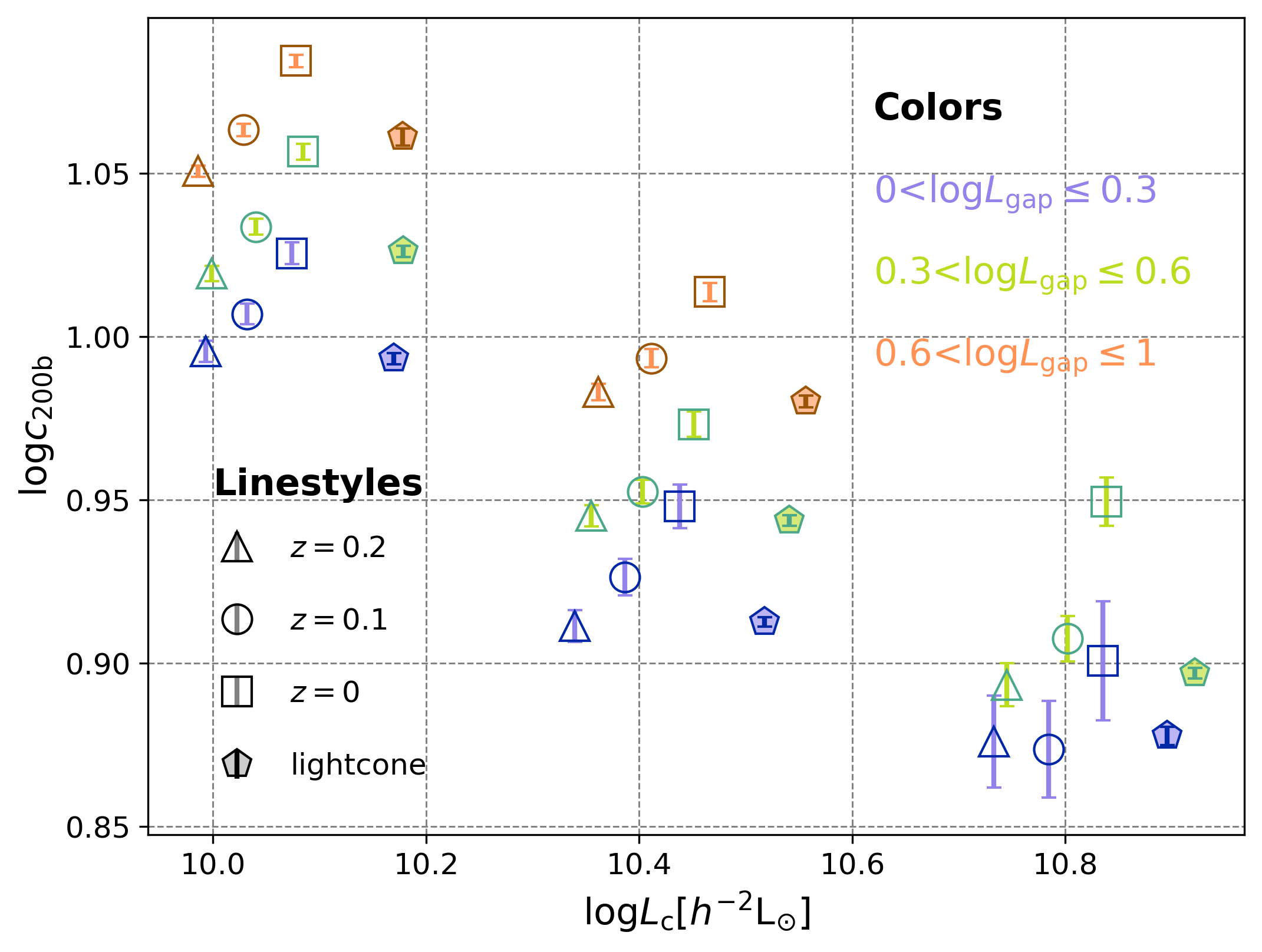}
    \caption{The impact of selection effects on the concentration - gap relation. Everything else are the same as in Figure~\ref{fig:logM_snapshots_lightcone}.}
    \label{fig:logc_snapshots_lightcone}
\end{figure}

In cosmological simulations, the data extracted from a snapshot represents galaxies with a complete luminosity distribution above the resolution limit within the simulation box. However, the real ICG sample is observed within a cone-shaped volume with a flux limit. The changing volume fractions with redshifts and the incompleteness of galaxy luminosity caused by the flux limit both contribute to the mean halo properties in the observational sample deviating from those in a simulation snapshot, or, a simple combination of multiple snapshots. To enable a fair comparison between observational results and theoretical predictions, it is essential to build lightcone samples at the theoretical end, where observational selection effects are incorporated into the simulation data. 

In this study, we compare two sets of simulation results with observations: one from the Millennium simulation, the database of which includes multiple lightcone samples for choice with different semi-analytical modeling (SAM) recipes of galaxies, and the other from TNG300, which doe not provide a lightcone sample. We demonstrate in this section how to construct a lightcone sample using TNG300 snapshots.

To construct the lightcone sample, we split the redshift distribution of the observational ICG sample as shown in Figure~\ref{fig:obs_z_dist} into 200 bins. For each very fine redshift bin, we calculate the number of ICGs within the bin and the lowest luminosity that can be observed based on the flux limit of $17.77$ and the median redshift of the bin. We then sample the same number of galaxy systems with the SBG above this luminosity threshold from the snapshot that is closest to the median redshift of the bin. By construction, this sample has the same redshift distribution and flux cut as the observational one.

Figure~\ref{fig:logM_snapshots_lightcone} shows the average relation between the logarithmic halo mass and the galaxy properties from three simulation snapshots and the lightcone sample. When calculating this relation, we select ICGs as central galaxies that are brighter than all satellite galaxies in the same friends-of-friends (FoF) halo. As illustrated by the three snapshots, systems with higher ICG luminosity are hosted by more massive halos, which is consistent with the stellar to halo mass relation (SHMR). At a given $L_\mathrm{c}$, systems with larger $L_\mathrm{gap}$ have lower halo mass. The mean $\log M_\mathrm{200b}$ increases as redshift decreases. 

Both the increase of halo mass with central luminosity and the gap dependence of halo mass in each $L_\mathrm{c}$ bin are preserved in the lightcone sample as in every snapshots. The mean $\log M_\mathrm{200b}$ in each $L_\mathrm{c}$ bin is around the middle of the three snapshots, which is intrinsic since the lightcone sample is composed of galaxy groups from all these redshifts. The gap dependence within $L_\mathrm{c}$ bins features slight difference from what is presented in snapshots, especially in the smallest two $L_\mathrm{c}$ bins. The difference is due to the flux cut. The observational flux limit cut the sample incomplete in each bin of $L_\mathrm{c}$ and $L_\mathrm{gap}$. Only systems with brighter $L_\mathrm{c}$ and smaller $L_\mathrm{gap}$, or, with brighter luminosity of the SBG, $L_s$, survive in the lightcone sample. These are the systems hosted by more massive halos, thus the mean $\log M_\mathrm{200b}$ is leveled up. Such increase of the mean $\log M_\mathrm{200b}$ is more significant at larger $L_\mathrm{gap}$ bins at a given $L_c$. Such bins have intrinsically lower $L_s$. Under an absolute luminosity limit, more ICGs in such bins are unobservable, thus making the lightcone sample more incomplete in these bins. In bins where the samples are more complete, the effect of the flux cut is not significant. The varying responses to the flux cut across gap bins causes the gap dependence in the lightcone sample to become tighter than those in snapshots.

Concentrations, on the other hand, present opposite trends with both $L_\mathrm{c}$ and $L_\mathrm{gap}$ at a given $L_\mathrm{c}$. As can be seen in Figure~\ref{fig:logc_snapshots_lightcone}, systems with higher $L_\mathrm{c}$ are embedded in halos with lower concentration, consistent with the expectation that more massive halos have lower concentrations \citep[see e.g.][]{Duffy_2008, klypin_halos_2011, prada_halo_2012, bhattacharya_dark_2013, diemer_universal_2014, ishiyama_uchuu_2021} . At a given $L_
\mathrm{c}$, larger $L_\mathrm{gap}$ systems have more concentrated halos, also consistent with theoretical expectations. The difference between the three snapshots shows that the mean $\log c_\mathrm{200b}$ in each bin of $L_\mathrm{c}$ and $L_\mathrm{gap}$ increases as redshift decreases. The lightcone sample represents a combination of the relations from the three snapshots due to the observational selection effect. 

\subsection{Comparing the gap dependence of halo properties with simulations}
\label{sec:compare with simulations}
\begin{figure}
    \includegraphics[width=\linewidth]{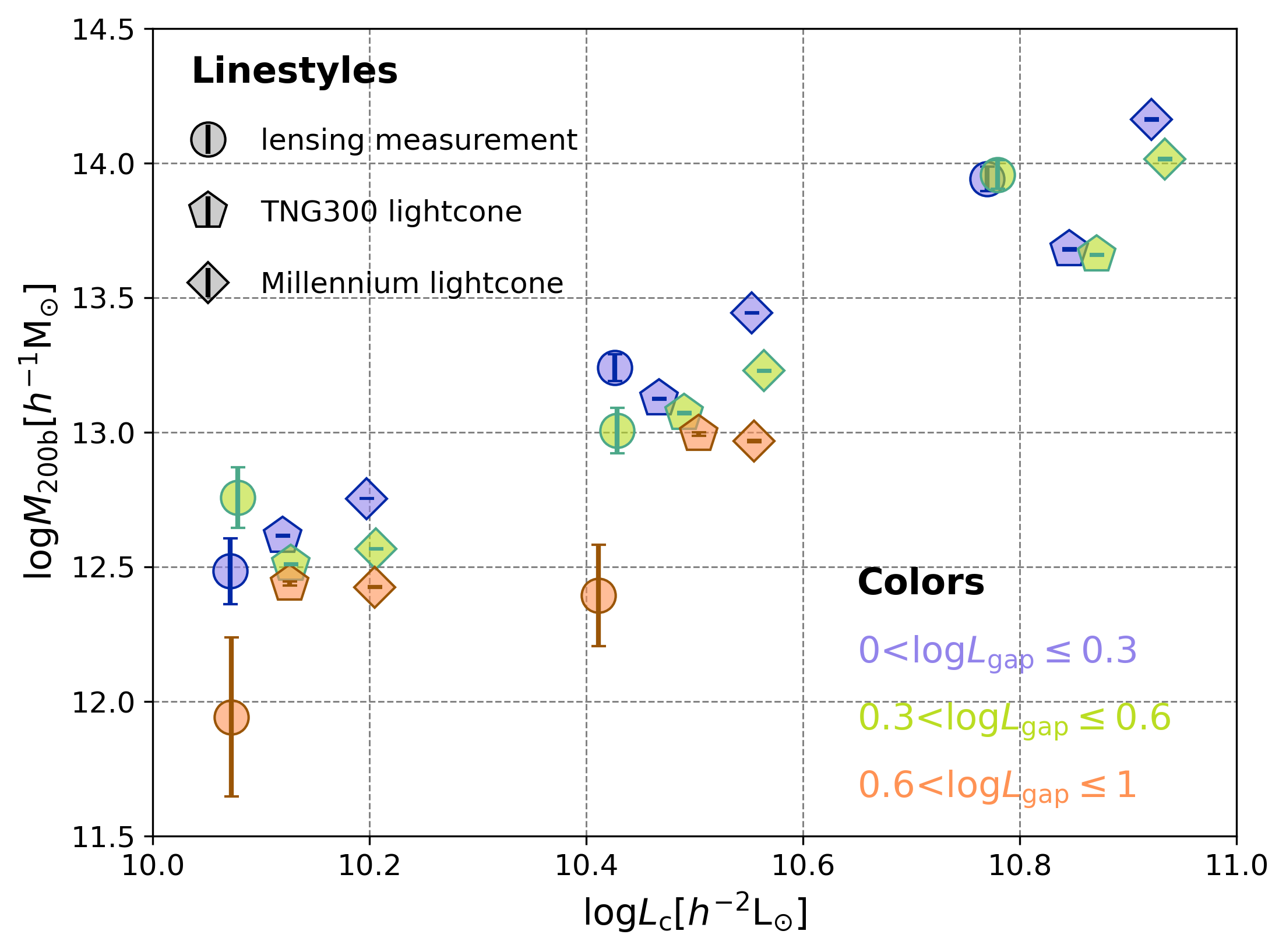}
        \caption{The halo mass - gap dependence in observations and simulations. The $y$ - coordinates of the empty circles represent best-fit $\log \hat{M}$ values from lensing ESD profiles (see Figure~\ref{fig:ESD-data}), calibrated to estimate the mean $\log M_\mathrm{200b}$ within each $L_\mathrm{c}-L_\mathrm{gap}$ bin. The $x$ - coordinates of the circles correspond to the mean luminosity of ICG in each bin. The error bars represent the $1\sigma$ uncertainty of the bias-calibrated estimators. Pentagons show the mean $\log M_\mathrm{200b}$ versus the mean $\log L_\mathrm{c}$ in a lightcone sample from TNG300, with errorbars representing the $1\sigma$ uncertainty of the sample mean $\log M_\mathrm{200b}$. Similarly, diamonds with error bars illustrate the mean $\log M_\mathrm{200b}$ versus the mean $\log L_\mathrm{c}$ and the $1\sigma$ uncertainty of the sample mean $\log M_\mathrm{200b}$ in the Henriques2012a~\citep{henriques_confronting_2012} lightcone sample of the Millennium simulation. Different colors denote results for different $L_\mathrm{gap}$ bins, as indicated by the legend. To more clearly display the results, we shift the $x$ - coordinates of the circles and diamonds $0.02$ dex to the left and $0.08$ dex to the right, respectively. }
    \label{fig:Mh_modelcomp}
\end{figure}
\begin{figure}
    \centering
    \includegraphics[width=\linewidth]{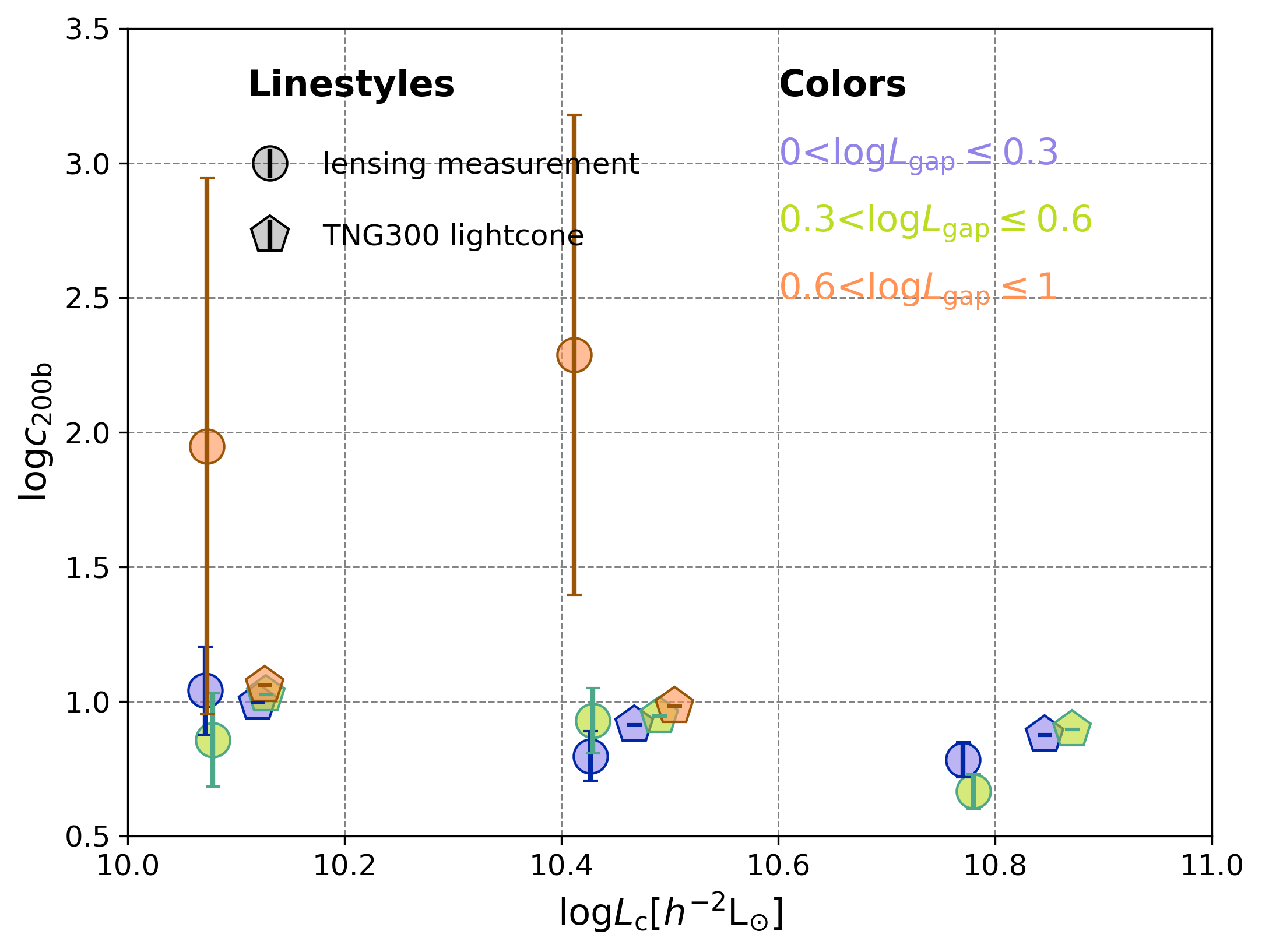}
    \caption{Similar to Figure~\ref{fig:Mh_modelcomp}, but showing the halo concentration - gap dependence. The data from the Millennium lightcone is missing because halo concentrations are not provided in the database.}
    \label{fig:c_modelcomp}
    \end{figure}
\begin{figure}
    \centering
    \includegraphics[width=\linewidth]{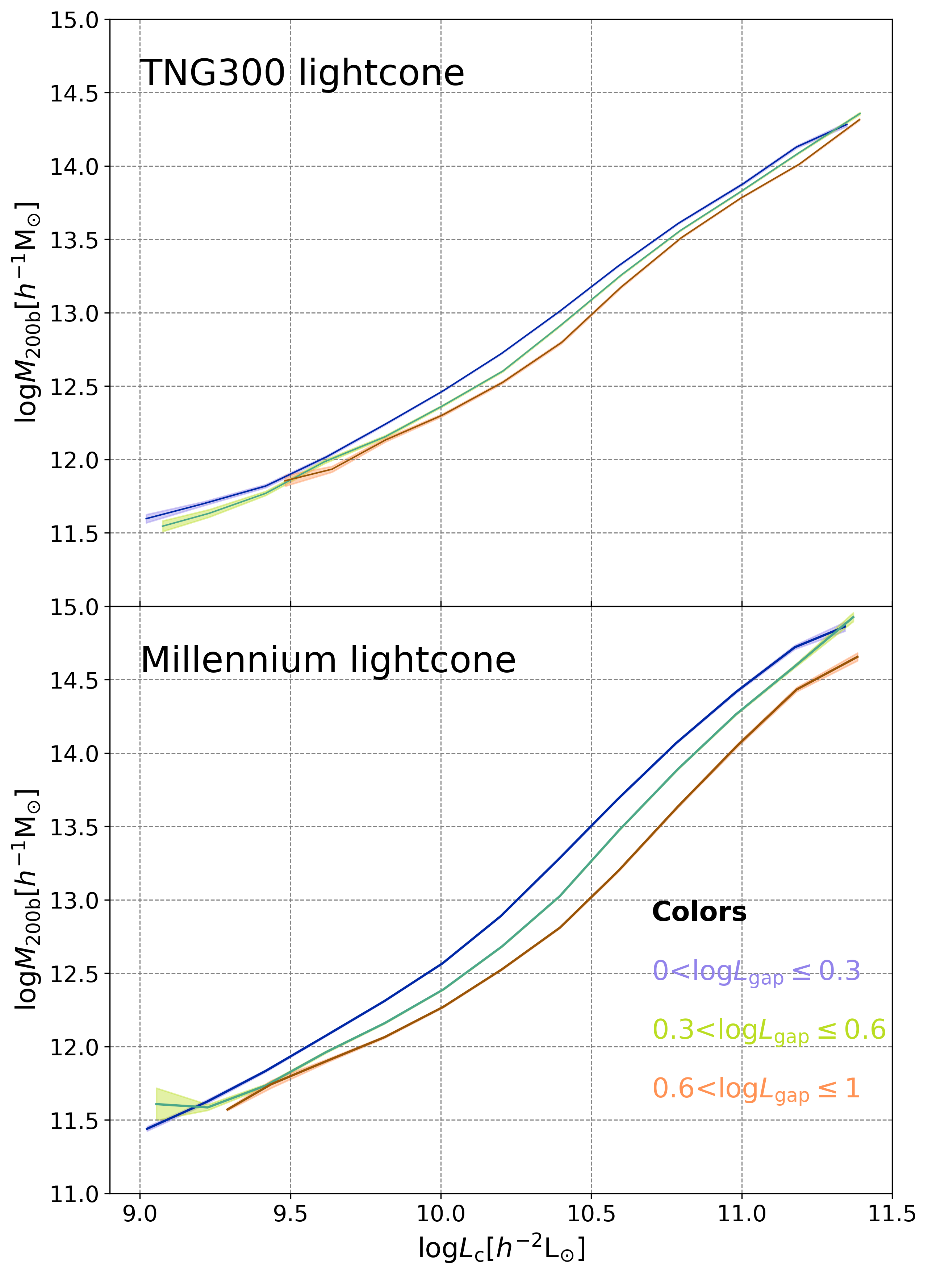}
    \caption{Halo mass dependence on the magnitude gap for ICGs within the $L_\mathrm{c}$ range of $10^{8.9}<L_\mathrm{c}[h^{-2}L_\odot]\leq 10^{11.5}$. The top and bottom panels show the results in the TNG300 and Millennium lightcone samples, respectively. Solid lines show the mean $\log M_\mathrm{200b}$, while shaded regions represent the $1\sigma$  uncertainty of the mean.
    Colors correspond to different gap bins as indicated by the legend.}
    \label{fig:logM_wider_Lc}
\end{figure}
\begin{figure}
    \centering
    \includegraphics[width=\linewidth]{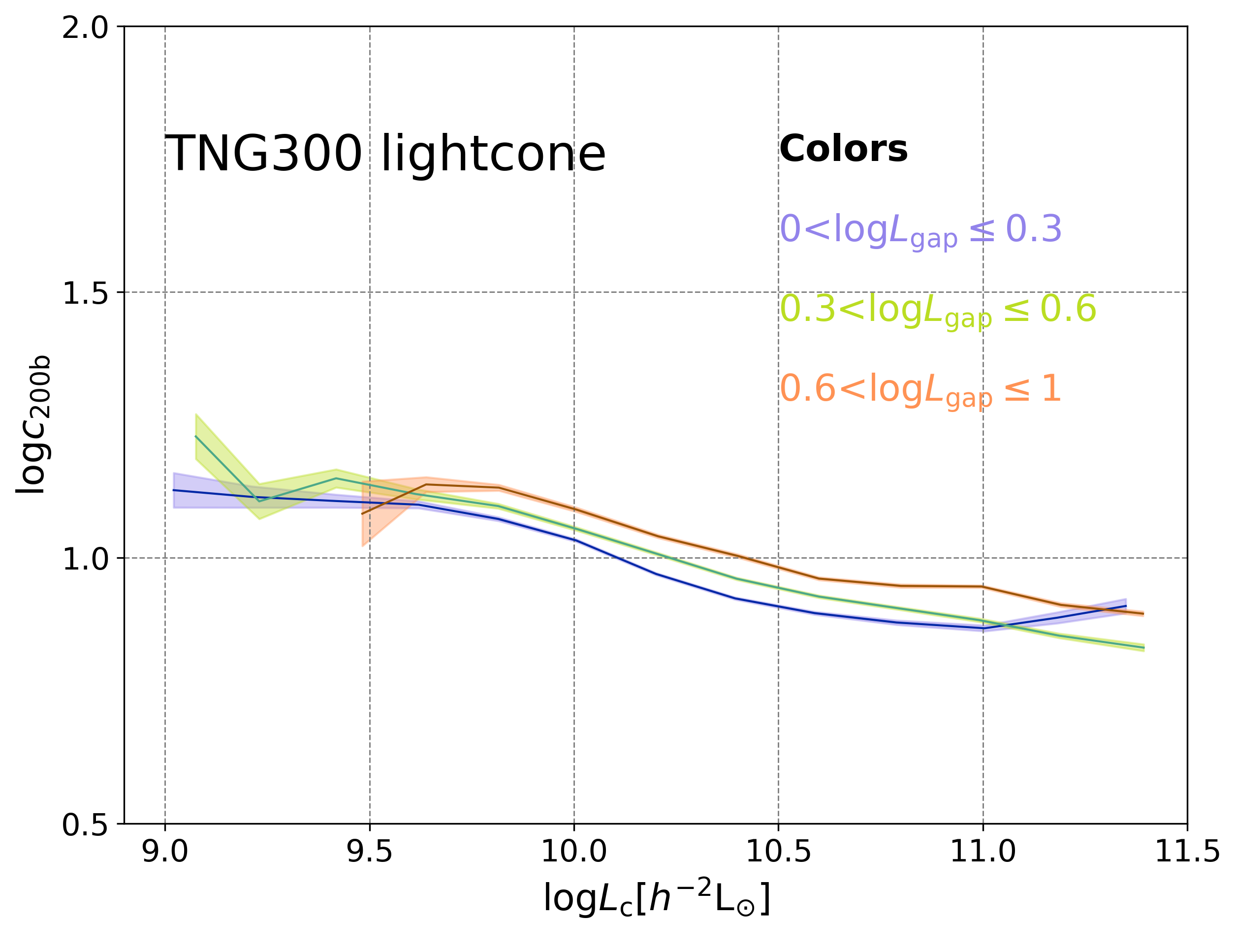}
    \caption{Similar to Figure~\ref{fig:logM_wider_Lc}, but showing the dependence of halo concentration on the magnitude gap for ICGs. Only the results in the TNG300 lightcone sample is shown, given that the Henriques2012a lightcone sample based on the Millennium simulation lacks the concentration information.}
    \label{fig:logc_wider_Lc}
\end{figure}

We now compare the dependence of halo properties on magnitude gaps in different simulations with our detection from g-g lensing, assessing variations of different models of galaxy formation and evolution. To ensure a fair comparison, we adjust all results to estimate the mean halo properties ($\log M_\mathrm{200b}$ and $\log c_\mathrm{200b}$) within each $L_\mathrm{c}-L_\mathrm{gap}$ bin, accounting for selection effects. For the observational case, we calibrate the lensing best fits to estimate the mean value in each $L_\mathrm{c}-L_\mathrm{gap}$ bin of the observational lens sample according to Equation~\ref{eq:M_bc} to~\ref{eq:var_c_bc} (see also Sections~\ref{sec:Mh_c_gap} and~\ref{sec: model est bias}). The simulation results come from two sources: the TNG300 lightcone sample, which we construct, and the Henriques2012a lightcone sample in the Millennium database. In the TNG lightcone, baryonic components are directly simulated, whereas the Henriques2012a lightcone relies on semi-analytical modeling (SAM) to account for baryonic processes. For more details on the TNG300 and Millennium simulations, as well as the Henriques2012a lightcone sample, we refer the readers to Section~\ref{sec:TNG300} and~\ref{sec:Millennium}. The procedures for constructing the TNG300 lightcone sample are described in Section~\ref{sec:model sel eff}.

The dependence of halo mass and concentration on the magnitude gap in the two simulation lightcones, along with the observational results, is shown in Figures~\ref{fig:Mh_modelcomp} and~\ref{fig:c_modelcomp}. The observational results are the same as the bias-calibrated ones in Figures~\ref{fig:logM_before_n_after_calibration} and~\ref{fig:logc_before_n_after_calibration}. 

Qualitatively, both simulation lightcones show trends of gap dependence that are broadly consistent with the observations. To make this point clearer, we show the dependence of halo mass and concentration on the magnitude gap in an extended $L_\mathrm{c}$ range in Figures~\ref{fig:logM_wider_Lc} and~\ref{fig:logc_wider_Lc}. The data come from the lightcone samples of both TNG300 and Millennium. The results reveal that the $M_\mathrm{200b}$ and $c_\mathrm{200b}-L_\mathrm{c}$ relations in a fixed $L_\mathrm{gap}$ bin follow approximate double power laws. The gap dependence is strongest at a certain value of $L_\mathrm{c}$ and weakens towards fainter and brighter $L_\mathrm{c}$ ends. This is a pattern that the lensing observations confirm in Figures~\ref{fig:Mh_modelcomp} and ~\ref{fig:c_modelcomp}. The Millennium lightcone displays a reversal trend of $\log M_\mathrm{200b}-L_\mathrm{gap}$ dependence for the smallest two $L_\mathrm{gap}$ bins in the very faint $L_\mathrm{c}$ end of around $\log L_\mathrm{c}[h^{-2}\mathrm{L}_\odot]=9.2$. For concentration, the TNG300 lightcone shows that the trend of $L_\mathrm{gap}$ dependence reverses from about $\log L_\mathrm{c}[h^{-2}\mathrm{L}_\odot]=11.1$ and maintains reversed at brighter ranges. These are also consistent with the `anomaly' we found in observations, though the specific $L_\mathrm{c}$ value at which the reversal occurs differs. 

Quantitatively, however, neither simulation fully matches the observed trends. In general, the TNG300 lightcone produces a weaker gap dependence than the weak lensing detection. The Millennium lightcone produces relatively stronger gap dependence, but still doesn't match the observational results in all $L_\mathrm{c}$ bins. This, together with the fact that the reversal trend occurs at different $L_\mathrm{c}$ between simulation ligtcones and the observation, suggests different turning points and distinct patterns of slope variation with $L_\mathrm{gap}$ in the double power law among the lensing detection and the two simulation lightcones. Such difference indicates deviations of the modeling of galaxy formation and evolution from the case in the real universe. 

\section{Discussion: The significance of halo mass and concentration dependence on magnitude gaps}\label{sec:significance}

To investigate the sensitivity of halo mass and concentration to the magnitude gap, we introduce three additional models for comparison. We refer to the model we adopted so far for all the results above as M1. In this model, both halo mass and halo concentration are allowed to vary for different $L_\mathrm{c}$ and $L_\mathrm{gap}$ bins. It has 16 free parameters in total: 2 free parameters ($M$ and $c$) for each of the 8 bins in $L_\mathrm{c}$ and $L_\mathrm{gap}$. Results for this model have been presented in Figures~\ref{fig:Mh_modelcomp} and~\ref{fig:c_modelcomp} above.

The three additional models are all less flexible than model M1 used so far. For model M2, we allow $M$ to change for different $L_\mathrm{c}$ and $L_\mathrm{gap}$ bins. On the other hand, $c$ is only allowed to change with $L_\mathrm{c}$. In other words, for a fixed bin of $L_\mathrm{c}$, the different $L_\mathrm{gap}$ bins are required to have the same best-fit halo concentration, based on their joint likelihood. Model M3 is similar, but we allow $c$ to change for different $L_\mathrm{c}$ and $L_\mathrm{gap}$ bins, whereas $M$ is only allowed to change with $L_\mathrm{c}$. Lastly, model M4 is the least flexible. For this model both $M$ and $c$ are only allowed to change with $L_\mathrm{c}$. We call M4 the null model given the absence of gap dependence. The definitions of these models are summarized in Table~\ref{tab:significance}.

The likelihood ratio test (LRT) is adopted to compare the three alternative models M1, M2 and M3 with respect to the null model M4. The likelihood ratio is calculated as $\Lambda_\mathrm{i}=2\ln(\mathcal{L}(\hat{\theta_\mathrm{i}})/\mathcal{L}(\hat{\theta_0}))$, where $\mathcal{L}(\hat{\theta_\mathrm{i}})$ and $\mathcal{L}(\theta_0)$ represent the likelihoods of the alternative models Mi (for i=1,2,3) and the null model M4, respectively, evaluated at their maximum likelihood estimates. 
Under the null hypothesis, $\Lambda$ asymptotically follows a $\chi^2$ distribution with degrees of freedom equal to the difference in the number of parameters between the alternative and null models \citep{wilks_large-sample_1938}. The $p$-value is then calculated as $p=1 - F_{\chi^2}(\Lambda; \text{df})$, where $F_{\chi^2}$ is the cumulative distribution function of the $\chi^2$ distribution. The $p$-value states the probability of observing a test statistic as extreme as $\Lambda$, assuming the null hypothesis is true. It represents how likely such a large test statistic could arise solely due to statistical variation under the null model. Thus, a smaller $p$-value indicates a higher significance of the alternative model.

The $p$-values corresponding to the likelihood ratio test statistics are listed in Table~\ref{tab:significance} for models M1, M2, and M3, respectively. These results indicate that M1 is the most statistically supported alternative model, providing the strongest evidence against the null model, followed by M2 and then M3. It suggests that allowing both halo mass and concentration to be dependent on the magnitude gap is the most data-favored among the four models, and the dependence of the magnitude gap on halo mass is stronger than its dependence on halo concentration. 

\begin{table}
    \centering
    \caption{The significances of models with or without the gap dependence in the halo parameters. The $p$-values indicate the probability that fitting with the assumed model is consistent with fitting statistical fluctuations in the null model. See details in Section~\ref{sec:significance}.}
\label{tab:significance}    
    \begin{tabular}{ccc}
    \hline\hline
      Models   &   Gap dependences&p-value\\\hline
         M1& 
     $M(L_\mathrm{c}, L_{\rm gap})$, $c(L_\mathrm{c}, L_{\rm gap})$&$1.8\times 10^{-6}$\\
 M2& $M(L_\mathrm{c}, L_{\rm gap})$, $c(L_\mathrm{c})$&$1.3\times 10^{-4}$\\
 M3& $M(L_\mathrm{c})$, $c(L_\mathrm{c}, L_{\rm gap})$&$4.6\times 10^{-1}$\\
 M4 (null model)& $M(L_\mathrm{c})$, $c(L_\mathrm{c})$&             ...\\
 \hline 
 \end{tabular}
\end{table}

\section{Conclusion}
\label{sec:conclusion}

We have investigated the dependence of halo mass and concentration on the central-satellite magnitude gap, with the halo properties measured directly through weak lensing. We select central galaxies as the brightest galaxy within a certain projected radius and line of sight velocity range according to the estimated virial size of its host halo. The magnitude gap is measured as that between the central galaxy and the second brightest galaxy within the same radius and velocity range. Applying the selections to a large sample of spectroscopic galaxies primarily from SDSS, we construct a sample of more than 30,000 central galaxies with measured magnitude gaps. We measure the stacked mass profiles of their host haloes in three central luminosity and two to three magnitude gap bins, to study the dependence of profile parameters on these galaxy properties. Our findings are compared with predictions from lightcone catalogs following the same selections in the TNG300 and the Millennium simulations. The key conclusions from our study are summarized below:

\begin{itemize}
\item We detect clear dependence of both host halo mass and concentration on the magnitude gap in g-g lensing. At a given central luminosity, systems with a larger magnitude gap generally have a lower halo mass and a higher concentration. The gap dependence is the most significant in the intermediate central luminosity bin in our sample with $L_{\rm c}\sim 10^{10.4} h^{-2} L_\odot$. 

\item In the lowest luminosity bin, the constraints are slightly weaker due to the larger observational error, and the gap dependences become non-monotonic, with the intermediate gap bin having the highest mass and lowest concentration. 

\item In the highest luminosity bin, no significant gap dependence is detected in either the halo mass or the concentration. This result is consistent with an interpretation that these galaxies reside in halos of $10^{14} h^{-1}\msun$ with the highest mass to light ratio, so that the mass increment brought in by merging satellites with lower mass to light ratios becomes less important.

\item Applying the lensing analysis pipeline to mock lensing catalogs built from simulations, we find that the best-fit halo mass and concentration parameters from our stacked lensing profiles using the PDF-symmetrization method is very close to the mean parameters of the stacked halos in logarithmic space, with negligible biases between the them.

\item The mass and concentration parameters for halos of a given central luminosity and magnitude gap both decrease towards higher redshift in the TNG simulation. The lightcone catalog combining the snapshots with observational selections qualitatively conserves the luminosity and gap dependence in the individual snapshots.

\item Results from both the TNG and the Millennium simulations show generally consistent gap dependence with observation. The gap dependence is the strongest in the intermediate luminosity range, and weakens towards both high and low luminosity ends. At the low luminosity end, the gap dependence of halo mass also starts to show non-monotonicity as in observations. This is a reflection of the double power-law shape of the halo mass-central luminosity relation, which takes on slightly different shapes in different gap bins and intersect at the low luminosity end. However, the intersection point generally occurs at a lower luminosity in simulations than in observations.

\item The observed gap dependence of halo mass is closer to that in the Millennium simulation, although the simulation tends to slightly over-predict the halo mass at the intermediate to high central luminosity. The gap dependences of both halo mass and concentration are quantitatively weaker than in observations.

    \item Fitting the stacked lensing profiles with and without gap dependences in mass and concentration, we find that a model allowing for the magnitude gap dependence on both halo mass and concentration has the highest statistical significance compared with models with a single or zero dependence. If only one parameter is allowed to depend on the magnitude gap, the likelihood ratio test indicates that the dependence of halo mass on magnitude gap is statistically more significant than that of concentration.
\end{itemize}

These results confirm that the magnitude gap can provide additional constraint on the halo profile besides the central luminosity. The qualitative differences in the measured gap dependences from simulations also open up new windows for diagnosing and improving the galaxy formation recipes in hydro simulations and semi-analytical models.

The ICG sample in our analysis provides reliable measurements for only eight data points, limiting the scope to a non-parametric analysis. However, with the advent of the ongoing and future large sky spectroscopic surveys such as DESI \citep{2016arXiv161100036D} and CSST \citep[e.g.,][]{Jiutian,CSSTmock}, we will be able to compile larger samples that extend to fainter luminosity limits. The expanded dataset will enable parametric modeling of the gap dependence of halo properties, facilitating more detailed comparisons between observational results and simulations. 

Data used in this paper can be shared upon request to the authors. The ESD profiles from lensing measurement, as well as the best-fit NFW halo parameters, are publicly available in Zenodo at \href{https://zenodo.org/records/15735889}{[DOI: 10.5281/zenodo.15735889]}.

\section*{Acknowledgments}

We are grateful to Johannes U. Lange for his guidance during the initial stages of this project, when we tried to do parametric fit of the gap dependence using the nested sampling Python package `Nautilus', which he developed. 

This work is supported by the National Key R\&D Program of China (2023YFA1605600,  2023YFA1605601, 2023YFA1607800, 2023YFA1607801, 
2023YFA1607804), the National Key Basic Research and Development Program of China (2023YFA1607800, 2023YFA1607802), NSFC (12022307, 12273021), the Fundamental Research Funds for the Central Universities, 111 project No. B20019, and Shanghai Natural Science Foundation, grant No.19ZR1466800. Additional support is provided by the Yangyang Development Fund. The computations of this work are carried on the Gravity supercomputer at the Department of Astronomy, Shanghai Jiao Tong University. 




\section{APPENDIX}

\subsection{Tests on the choice of projected radial range for NFW model profile fitting}
\label{sec:test radial choice}

\begin{figure*}
    \centering
    \includegraphics[width=0.7\textwidth]{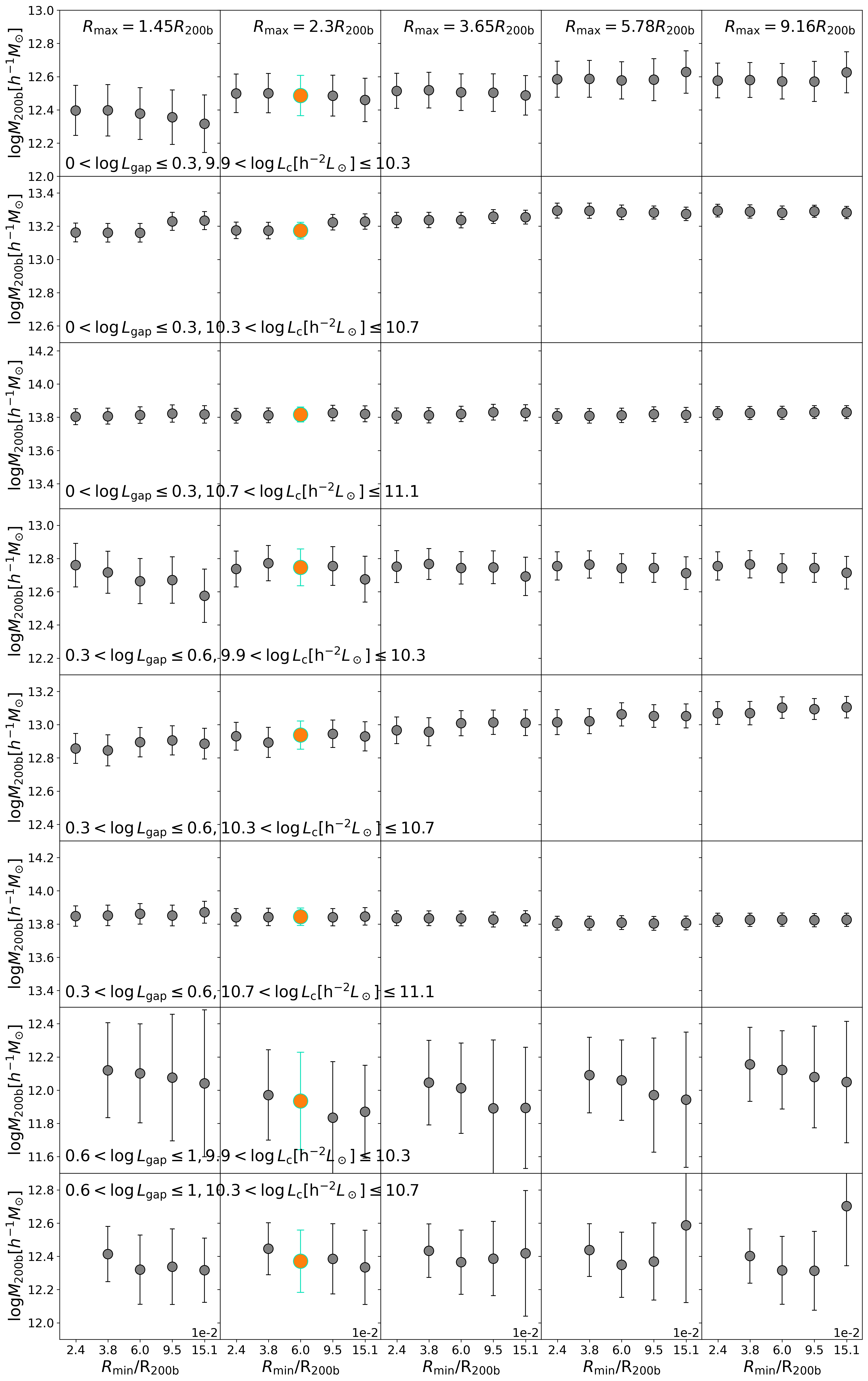}
    \caption{The best-fit (maximum-likelihood) host halo mass based on different choices of the projected radial ranges. Each row represents a $(L_\mathrm{c},L_\mathrm{gap})$ bin (see the text in the left most panel of each row). In the same row, each panel refers to a given choice of the outer radius ($R_\mathrm{max}$), denoted by the text in each panel of the first row. In a given panel, different data points are the best recovered halo mass with different choices of the inner radius ($R_\mathrm{min}$), as indicated by the $x$-axis. The colored data points mark the best fits under our default fitting range choice of $R_\mathrm{min} = 0.06R_\mathrm{200b}$ and $R_\mathrm{max}=2.3R_\mathrm{200b}$.}
    \label{fig:M_r}
\end{figure*}

In this appendix we test the sensitivity of our best recovered host halo mass and concentration parameters on the different choices of radial range in projection (from $R_\mathrm{min}$ to $R_\mathrm{max}$). We try five different choices of $R_\mathrm{max}$ as well as five different choices of $R_\mathrm{min}$ at fixed $R_\mathrm{max}$. Over the projected radial range between $R_\mathrm{min}$ and $R_\mathrm{max}$, we fit projected NFW profiles to get the best-fit halo properties, with the results shown in Figure~\ref{fig:M_r}. Each row refers to a given bin of $L_\mathrm{c}$ and $L_\mathrm{gap}$. Our default choice of $R_\mathrm{min}=0.06R_{200b}$ and $R_\mathrm{max}=2.3R_{200b}$ corresponds to the third data point in the second panel from the left in each row. For panels in a given row, the chosen $R_\mathrm{max}$ increases from the left to right panels. We can see that the best-fit halo mass increases from $R_\mathrm{max}=2.3R_{200b}$ to $R_\mathrm{max}=9.16R_{200b}$ in a few rows, which is likely associated with the two-halo term. For measurements in the same panel, the best-fit halo mass slightly decreases with the decrease in $R_\mathrm{min}$, which is likely due to the off-centering effect. However, the trends are not significant compared with the measurement errors, especially for the two left most panels in each row, indicating our results are not sensitive to the chosen $R_\mathrm{min}$ and $R_\mathrm{max}$ when they are varying within a reasonable range.



\bibliographystyle{aasjournalv7}
\bibliography{gap_lensing_ApJ} 

\end{CJK*}
\end{document}